\documentclass[12pt,preprint]{aastex}

\newcommand\diamondrule{\line{$\m@th
   \leaders\hrule\hfill\rlap{$\m@th\bracerd\braceld$}
   \braceru\bracelu\leaders\hrule\hfill$}}

\newcommand\smallneg{\kern-.0800em}
\newcommand\negskip{\kern-.5em}

\newcommand\lsim{\rlap{\raise.4ex\hbox{$<$}}\lower.55ex\hbox{$\sim$}\,}
\newcommand\gsim{\rlap{\raise.4ex\hbox{$>$}}\lower.55ex\hbox{$\sim$}\,}
\newcommand\implies{{\bf=\kern-0.45em>}}

\newcommand\unit{\,\rm}

\newcommand\kms{\rm\, km\cdot s^{-1}}
\newcommand\Kkms{\rm\, K\cdot\kms}

\newcommand\um{\unit\mu m}

\newcommand\CO{\rm {}^{12}\smallneg CO}

\newcommand\cO{\rm {}^{13}\smallneg CO}

\newcommand\cObf{\bf {}^{13}\smallneg CO}

\newcommand\cOit{{}^{13}\smallneg\it CO}

\newcommand\Jone{\rm J=1\rightarrow 0}

\newcommand\COone{\CO\ \Jone}

\newcommand\cOone{\cO\ \Jone}

\newcommand\nH{\rm n(H_2)}

\newcommand\Tk{\rm T_{{}_K}}

\newcommand\Tr{\rm T_{{}_R}}

\newcommand\Trz{\rm T_{{}_{R0}}}
\newcommand\Trone{\rm T_{{}_{R1}}}

\newcommand\ckms{\unit cm^{-2}\cdot (km\cdot s^{-1})^{-1}}

\newcommand\Td{\rm T_{d}}
\newcommand\Tdz{\rm T_{d0}}
\newcommand\Tdo{\rm T_{d1}}
\newcommand\Tdc{\rm T_{dc}}
\newcommand\DT{\rm \Delta T}
\newcommand\rd{\rm r_{_{240}}}

\newcommand\Ia{\rm I_\nu(140\um)}
\newcommand\Ib{\rm I_\nu(240\um)}
\newcommand\Ic{\rm I(\cO)}
\newcommand\sga{\rm \sigma(140\um)}
\newcommand\sgb{\rm \sigma(240\um)}
\newcommand\sgc{\rm \sigma(\cO)}
\newcommand\MJsr{\unit MJy\cdot sr^{-1}}
\newcommand\MJkk{\unit MJy\cdot sr^{-1}\cdot (K\cdot km\cdot s^{-1})^{-1}}
\newcommand\NHd{\rm N_d(H)}
\newcommand\NHdo{\rm N_{d1}(H)}

\newcommand\Nnth{\rm N_{13}(H)}

\newcommand\Dv{\rm\Delta v}
\newcommand\Dvc{\rm\Delta v_c}
\newcommand\nvco{\rm {N_{c1}\over\Dvc}}
\newcommand\nvcz{\rm {N_{c0}\over\Dvc}}
\newcommand\nvtco{\rm {N_{c1}(\cO)\over\Dvc}}
\newcommand\nvtcz{\rm {N_{c0}(\cO)\over\Dvc}}

\newcommand\NDv{\rm N(\cO)/\Delta v}

\newcommand\degree{\rlap{$^\circ$}\kern.06em} 
\newcommand{\mymail}{wwall@inaoep.mx}

\slugcomment{Version --  23/Jan/2006}

\shorttitle{Diagnostic of Dust/Molecular Gas Conditions}
\shortauthors{Wall}

\begin{document}


\title{Comparison of $\cObf$ Line and Far-Infrared Continuum\\
	Emission as a Diagnostic of Dust and Molecular Gas\\
	Physical Conditions:\\
	II. The Simulations: Testing the Method}


\author{W. F. Wall}
\affil{Instituto Nacional de Astrof\'{\i}sica, \'Optica, y Electr\'onica,
Apdo. Postal 51 y 216, Puebla, Pue., M\'exico}
\email{\mymail}

%%\and

%%\author{W. T. Reach}
%%\affil{California Institute of Technology, IPAC/SSC, Pasadena MS 220-6, California}
%%\email{reach@ipac.caltech.edu}

%\author{Tomoo Nagahama and Yasuo Fukui}
%\affil{Department of Astrophysics, Nagoya University, Chikusa-ku, Nagoya 464-8602, Japan}
%\email{nagahama@a.phys.nagoya-u.ac.jp, fukui@a.phys.nagoya-u.ac.jp}

%\and

%\author{R. J. Hanisch\altaffilmark{5}}
%\affil{Space Telescope Science Institute, Baltimore, MD 21218}


\begin{abstract}

The reliability of modeling the far-IR continuum to $\cOone$ spectral line ratios 
applied to the Orion clouds \citep{W05} is tested by applying the models to simulated 
data.  The two-component models are found to give the dust-gas temperature difference, $\DT$, 
to within 1 or 2$\,$K.  However, other parameters like the column density per velocity
interval and the gas density can be wrong by an order of magnitude or more.  In 
particular, the density can be systematically underestimated by an order of magnitude
or more.  The overall mass of the clouds is estimated correctly to within a few 
percent.

The one-component models estimate the column density per velocity interval and density
within factors of 2 or 3, but their estimates of $\DT$ can be wrong by 20$\,$K.  They
also underestimate the mass of the clouds by 40-50\%. 

These results may permit us to reliably constrain estimates of the Orion clouds' physical
parameters, based on the real observations of the far-IR continuum and $\cOone$ spectral
line.  Nevertheless, other systematics must be treated first.  These include the effects of
background/foreground subtraction, effects of the HI component of the ISM, and others.
These will be discussed in a future paper \citep{W05a}.   

\end{abstract}




\section{Introduction\label{sec1}}

Paper~1 \citep{W05} examined the ability of the FIR-continuum to $\cOone$
line intensity ratio to diagnose dust and molecular gas physical conditions.
Specifically, the {\it COBE/DIRBE\/} 140$\um$ and 240$\um$ continuum data
\citep[see][]{dirbex} were compared with the Nagoya 4-m $\cOone$ spectral line 
data for the Orion~A \citep{Nagahama98} and B molecular clouds.  The $\Ib/\Ic$ ratio, 
or $\rd$, was plotted against the 140$\um$/240$\um$ dust color temperature, or $\Tdc$, 
for the high signal-to-noise positions ($\ge 5-\sigma$ for 140$\um$, 240$\um$, and 
$\cOone$) in the Orion clouds.  This plot was modeled with LTE and LVG, one-component 
models and LVG, two-component models; the two-component models fit the data better than 
the one-component models at the 99.9\% confidence level.  Tables~1, 2, and 3 of Paper~I 
list the resultant parameter values of the two-component model fits.  The most noteworthy 
result is that the two-component models demand the dust-gas temperature difference, $\DT$, 
to be zero within $\pm 1$ or 2$\,$K. (Note that in the case of the two-component,
two-subsample models, the $\Tdc\ge 20\,$K subsample still yields $\DT = 0\pm 1K$
{\it if\/} a two-component model is fitted to that subsample.  The listed results
in Table~2 of Paper~I are those of the one-component model fitted to the 
$\Tdc\ge 20\,$K subsample.)  This result has important consequences that were
briefly mentioned in Paper~I and will be discussed in detail in Paper~III \citep{W05a}.
Consequently, the reliability of the derived $\DT$ must be tested.

In all of the modeling mentioned in Paper~I, the systematic uncertainties of the
derived parameter values were evaluated by applying scale factors to the data.  
These systematic uncertainties are related to uncertainties in the calibration and 
in certain assumptions, such as the dust optical depth to gas column density ratio.  
The combined effect of these uncertainties was estimated to be $\pm 40\%$.  Accordingly, 
scale factors that varied from 0.6 to 1.4 were applied to the data to see how strongly
the resultant parameter values would change.  Also, the starting search grid for the
two-component models was slightly shifted and re-run.  The magnitudes of the changes
in the results provided another test of the systematic uncertainties in the parameter
values.  These two tests gave similar estimates of the systematic uncertainties. 
These systematic uncertainties are demonstrated in Figure~21 of Paper~I, which shows
that the column densities per velocity interval and densities of both components are
uncertain by factors of a few or by more than an order of magnitude.  (These uncertainties
are orders of magnitude larger than the formal uncertainties obtained from the model fits.
Accordingly, the latter uncertainties can be ignored.) 

While the abovementioned tests provide rough estimates of the reliability of the 
results, they do {\it not\/} measure any biases inherent in the method.  In other
words, the range of possible parameter values that result from the modeling and
from the tests {\it may not even include the ``true'' or correct value.\/}  And we
cannot know that these ranges are indicative of the correct values,  because we cannot 
know the correct values in the first place.  This is in stark contrast to using simulated 
data.  With simulated data, the true, or input, values can be compared with the resultant
values from the model fits.  The tests that were applied to modeling the actual observed
data can be repeated on the modeling of the simulated data.  Biases or shortcomings in
the modeling technique are then clearly seen.  In the following section and its subsections, 
the creation of the simulated data and the results of modeling these data is 
described. 

Other systematics are not discussed in the current paper, but are left to Paper~III.  These 
are the systematic effects that result when the models do not properly characterize the 
contributions of other phases of the ISM, such as from HI and its dust or from some
large-scale foreground/background emission, or when they adopt an improper value of some 
more basic physical parameter, such as the far-IR spectral emissivity index, $\beta$.

\section{The Simulations\label{ssec37}}

    To better understand the strengths and weaknesses of determining gas and
dust physical conditions using the ratio of the FIR continuum to the $\cOone$
line, simulated data were created.  The simulations assumed that the real clouds
are composed of two components: a component 0 and a component 1.  The former
has constant physical conditions; i.e., they do not vary from one line of sight
to another.  The latter also has constant physical conditions, except for the
dust and gas temperatures (i.e. $\Td$ and $\Tk$).  The component-1 temperatures 
vary from line of sight to line of sight, but maintain a constant dust/gas 
temperature difference, $\DT\equiv\Td - \Tk$. The simulations started with a 
map of beam-averaged column densities (i.e., column densities that are averaged 
over $\sim 1^\circ$ scales) and component-1 dust temperatures. $\Tdo$.  Model 
parameters were specified for two subsamples and two
components (see Table~\ref{tbl-7} for details).  The two subsamples were the 
$\Tdo < 20\,$K points and the $\Tdo\geq 20\,$K points.  This 
is not exactly the same as using $\Tdc=20\,$K (where $\Tdc$ is the 140$\um$/$240\um$
color temperature) as the boundary (as was done in Paper~I), but, since 
$\Tdz=18\,$K and since the column density of component 1 within each velocity 
interval, i.e. $\nvco$, is factors of 4 to 10 larger than the corresponding
component-0 quantity, $\nvcz$ (see Table~\ref{tbl-7}), component~1 dominates 
the emission near the $\Tdo=20\,$K boundary by roughly an order of magnitude.  
Consequently, $\Tdc=20\,$K is equivalent to $\Tdo=20\,$K for all practical 
purposes.  The model intensity maps were then generated using the procedure below:
\begin{enumerate}
\item The map of $\Tdo$ values determined whether a given pixel
belonged to subsample$\,$1 or subsample$\,$2.
\item The subsample to which a pixel belongs then dictated which
model parameter values belonged to that pixel.  Using these
values in equation~(28) of Paper~I gave the area filling factor within a clump
velocity width, or the $c_1$ value, for that pixel.  The
observed velocity width, $\Dv$, adopted was 2$\kms$, which is the 
the actual observed velocity width in the Orion clouds in the $\cOone$ line
on the scale of $1^\circ$.  Nevertheless, the 
expressions that give the observed intensities (i.e., 27 and
29 of Paper~I) are actually independent of $\Dv$.  $\Dv$ only determines the 
filling factor, $c_1$. 
\item Equations (20), (27), (31), (32), and 
(29) of Paper~I then gave the $\Ia$, $\Ib$, and $\Ic$ intensities 
observable from that pixel.  In addition, the color corrections for bands 9 and
10 of {\it COBE}/{\it DIRBE\/} converted the $\Ia$ and $\Ib$ values to those
observable in the {\it DIRBE\/} bands. 
\item The intensities, $\Ia$, $\Ib$, and $\Ic$, then specified 
the uncertainties in those intensities, $\sga$, $\sgb$, and $\sgc$, based on 
the prescriptions described below and based on the 
observed data.  These uncertainties for all the pixels represent the 
$\sigma$ maps.
\item For the given pixel, a random number generator with a normally 
distributed probability of outputs with a mean of zero and an rms
dispersion of unity generated noise values in the three 
wavelength bands.  The noise value for each band was scaled by the 
$\sigma$ for that pixel and for that band (i.e. $\sga$, $\sgb$, or $\sgc$).  
These noise values for all the pixels represent the noise maps. 
\item The noise maps were then added to the noise-free intensity maps to
produce the final simulated maps.
\end{enumerate}

\begin{figure}
\epsscale{0.6}
\plotone{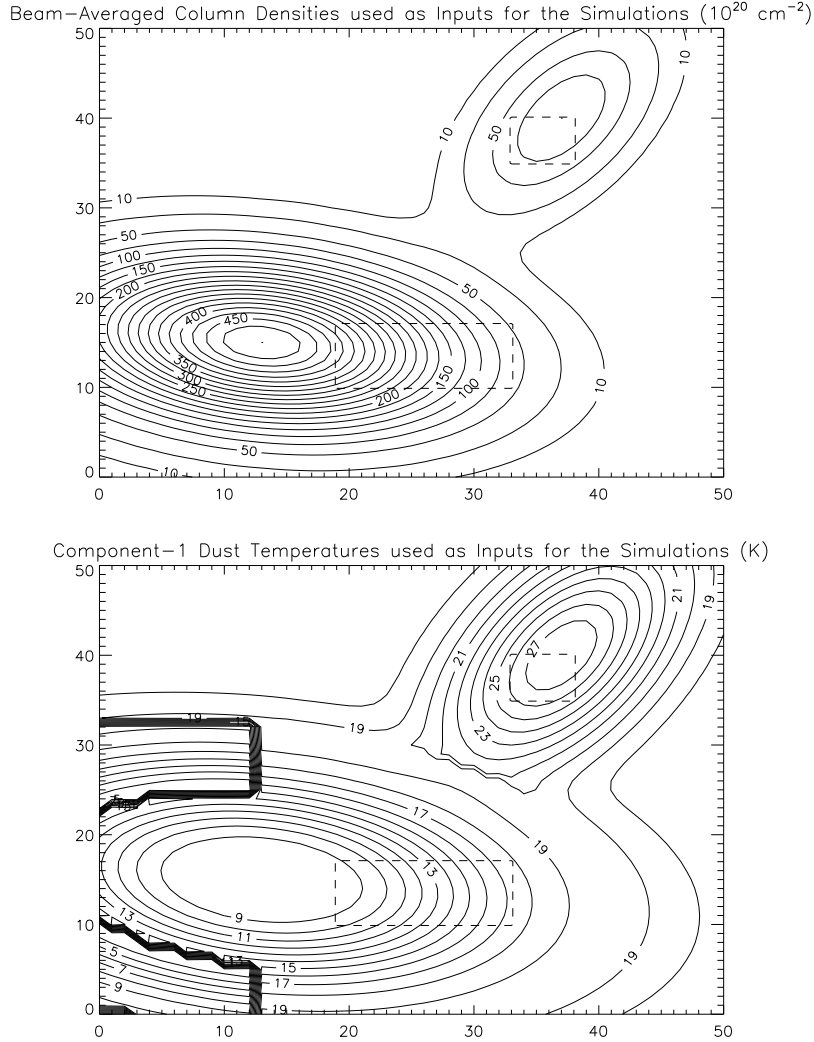}
\caption{The above panels contain contour maps of the inputs for the simulations.
The upper panel is the map of beam-averaged column densities in units of 
$10^{20}\ H\ nuclei\cdot\unit cm^{-2}$.  The lower panel is the map of component-1
dust temperatures in Kelvins.  Notice that while the column density map
consists of two peaks, the temperature has one peak, which coincides with
the column density peak in the upper right, and one depression, which coincides
with the column density peak in the lower left.  The temperature map also has extra
low values (between 3 and 12$\,$K) within the boundary of the closely spaced 
contours that appear on the left side of the map at the bottom and near the middle 
of the left side.  The dashed rectangles illustrate the positions of the patches 
that have the low noise values in the $\Ic$ map. 
\label{fig36}}
\end{figure}

    The noise prescriptions mentioned above are based on the $3\times 3$ 
smoothed maps of the real observations.  The uncertainties in these maps had 
approximately the following behavior:
\begin{eqnarray}
\sga &=& \left\{
\begin{array}{ll}
2\MJsr, & {\rm\quad for\ } \Ia\leq 60\MJsr\\
0.03\,\Ia, & {\rm\quad for\ } \Ia > 60\MJsr\\
\end{array}
\right.
\label{mr43}\\    
\sgb &=& \left\{
\begin{array}{ll}
0.5\MJsr, & {\rm\quad for\ } \Ib\leq 50\MJsr\\
0.01\,\Ia, & {\rm\quad for\ } \Ib > 50\MJsr\\
\end{array}
\right.
\label{mr44}\\
\sgc &=& \left\{
\begin{array}{ll}
0.05\Kkms, & {\rm outside\ Orion\,A\ Field}\\
0.005\Kkms, & {\rm inside\ Orion\,A\ Field}\\
\end{array}
\right.
\label{mr45}
\end{eqnarray} 
It should be mentioned that the sigma levels for the simulated 240$\um$ and $\cO$ maps
are actually half of those of the actual observed maps.  This reduction of the sigma
levels in the simulated 240$\um$ and $\cO$ maps was done to ensure a sufficient number
of high-sigma points.  To generate the
$\sigma$ map for $\Ic$, a portion of the map area
that would represent the lower-noise subfield within Orion$\,$A Field was chosen.  
The simulated maps were chosen to be 51$\,$pixels~$\times$~51$\,$pixels, a total 
of 2601 pixels and similar to that of the Orion fields: 2609.  The area 
designated to have the lower noise of the
Orion$\,$A Field consisted of two separate rectangular patches with a total of 
156 pixels. One of the patches included a peak in the input column density map 
and the other patch included areas of low column density (see Figure~\ref{fig36}).  
The patch with the column density peak also had a peak in the component-1 
dust temperature.  This was consistent with the actual observations.

     Now the input column density and component-1 dust temperature maps must 
be specified.  These maps are depicted in Figure~\ref{fig36}. The 
maximum column density was chosen to be roughly the same as that of
the observations (i.e., the two-component models): 
$5\times 10^{22}\ H\ nuclei\cdot\unit cm^{-2}$.  The column density map has 
two elliptical gaussians: one with a low peak that crudely
represents the Orion$\,$Nebula Field and one with a high peak that crudely
represents the main body of the Orion$\,$A molecular cloud.  In the 
Orion$\,$Nebula field, the dust temperature rises with rising column density.
Consequently, the component-1 dust temperature map has
an elliptical gaussian peak corresponding to the low peak in the
column density map.  In the main body of the Orion$\,$A cloud, however, the
dust temperature declines with increasing column density.  Therefore, the 
temperature map has an elliptical gaussian {\it valley\/}
corresponding to the high peak in the column density map.  The component-1
temperatures range from 3 to 28$\,$K.  To ensure that a small
minority of the pixels had sufficiently low temperature values, these values
were placed in two patches on the left edge of the map (see lower panel of
Figure~\ref{fig36}).  The procedure above was then implemented using
the parameter values in the first two columns of Table~\ref{tbl-7} to
yield the simulated maps. 

\begin{figure}
\epsscale{0.7}
\plotone{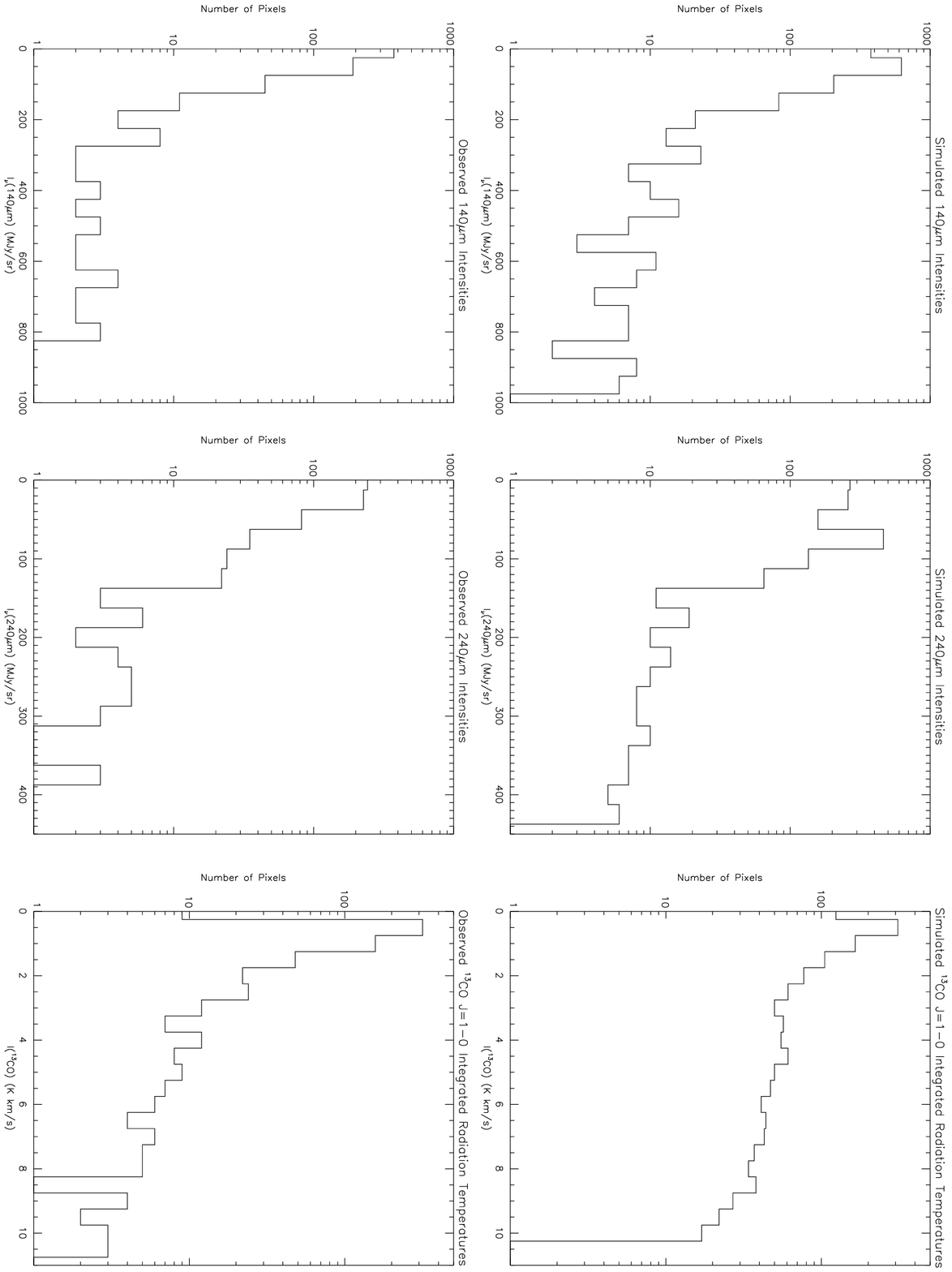}
\caption{Histograms of the simulated and observed intensities for the 140$\um$
and 240$\um$ continuum and the $\cOone$ line are shown.  The upper panels have
the histograms of the simulated data and the lower panels have the histograms 
of the observed data.  In all of the panels, only those pixels with 
intensities above the 5-$\sigma$ level in $\Ia$, $\Ib$, $\Ic$ {\it 
simultaneously\/} are represented in the histograms.  This corresponds to
a total of 1465 pixels in the simulated maps and 674 pixels in the observed 
maps.  
\label{fig37}}
\end{figure}

\begin{figure}
\epsscale{0.7}
\plotone{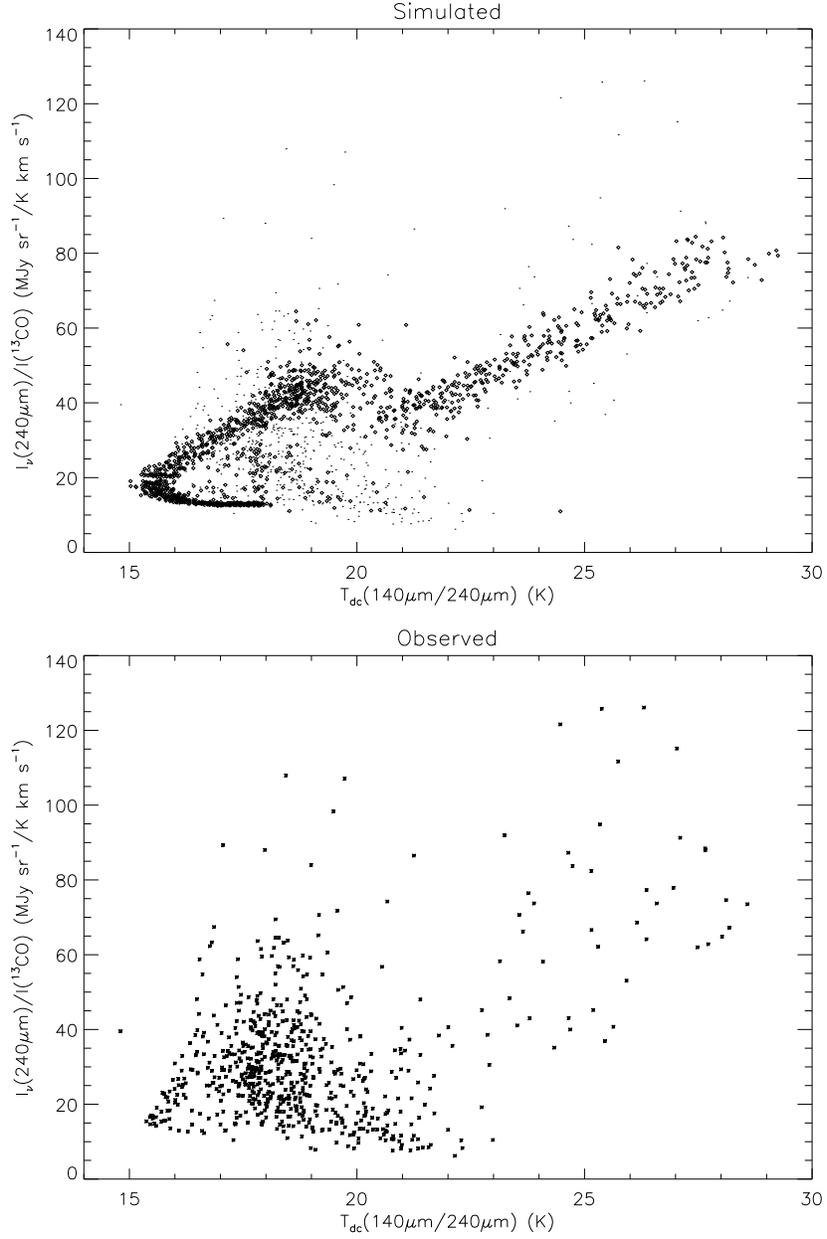}
\caption{Plots of $\rd$ versus the 140$\um$/240$\um$ color temperature are
shown for the simulations and for the observations.  The upper panel is
the plot for the simulated data and the lower panel is for the
observed data.  The error bars are omitted for clarity.  The panels only 
include those pixels with intensities above the 5-$\sigma$ level in $\Ia$, 
$\Ib$, $\Ic$ {\it simultaneously\/}. 
\label{fig38}}
\end{figure}

\begin{figure}
\epsscale{0.7}
\plotone{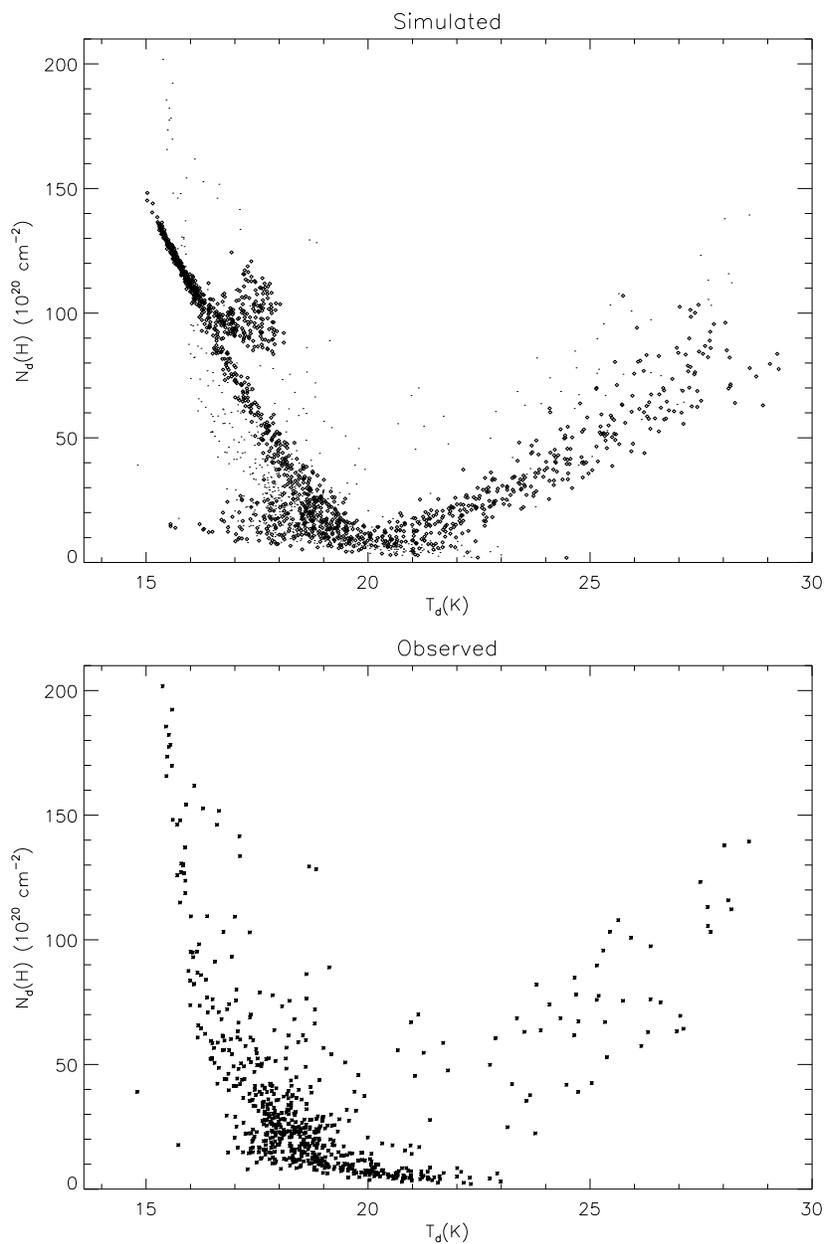}
\caption{Plots of continuum-derived gas column densities, $\NHd$, for the 
one-component case versus the dust temperature are given 
for the simulations and for the observations.  The column densities are in 
units of $10^{20}\ H\ nuclei\cdot\unit cm^{-2}$.  The upper panel is the plot 
for the simulated data and the lower panel is for the observed data.  
The panels include omit the error bars for clarity.    The panels only 
include those pixels with intensities above the 5-$\sigma$ level in $\Ia$, 
$\Ib$, $\Ic$ {\it simultaneously\/}.  
\label{fig39}}
\end{figure}

\begin{figure}
\epsscale{0.7}
\plotone{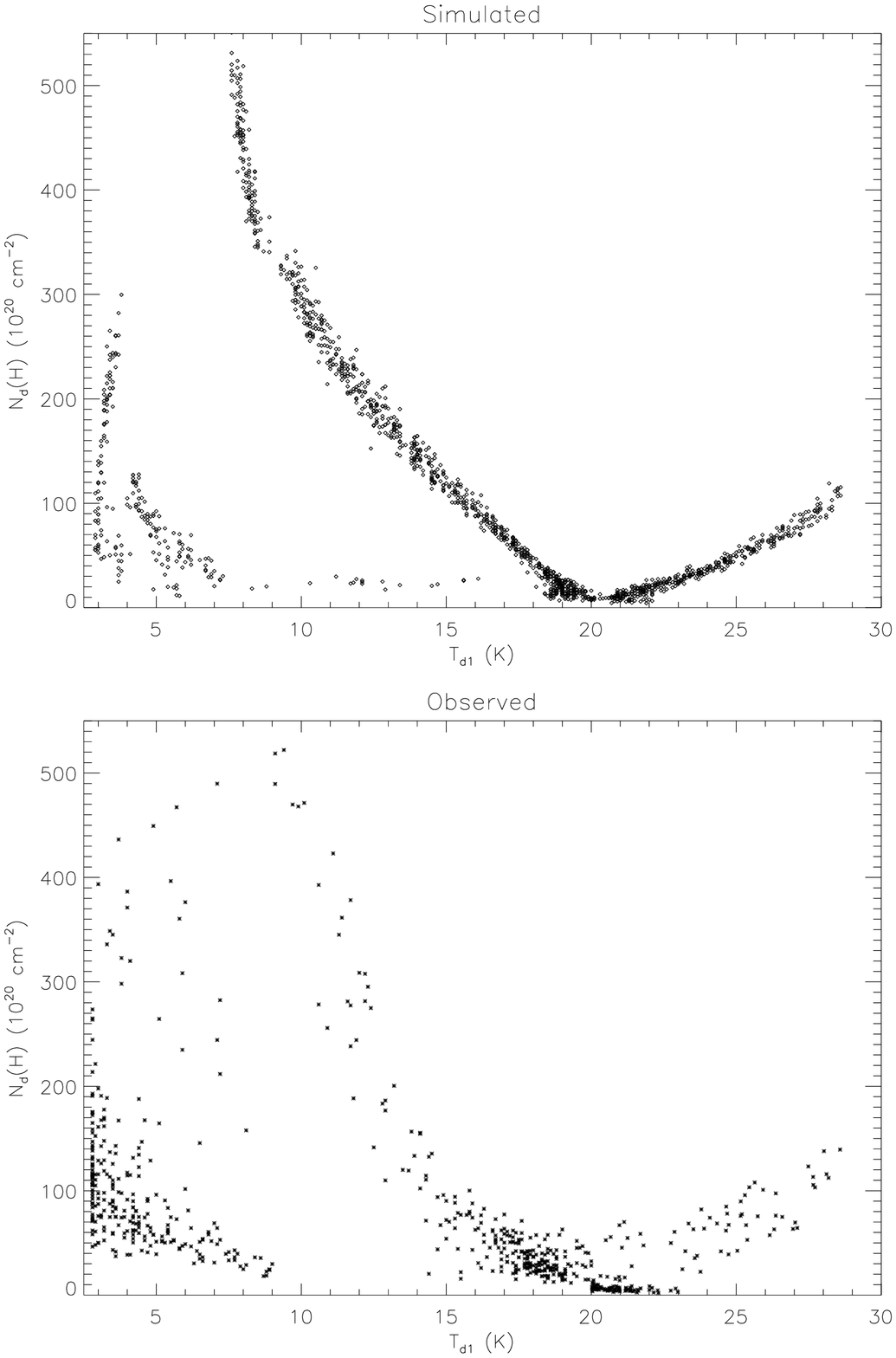}
\caption{Plots of continuum-derived gas column densities, $\NHd$, for the 
two-component, two-subsample case versus the component-1 dust temperature 
are shown for the simulations and for the observations.  The column densities 
are in units of $10^{20}\ H\ nuclei\cdot\unit cm^{-2}$.  The upper panel is the 
plot for the simulated data and the lower panel is for the observed 
data.  The panels omit the error bars for clarity. The panels only include 
those pixels with intensities above the 5-$\sigma$ level in $\Ia$, $\Ib$, $\Ic$ 
{\it simultaneously\/}.   
\label{fig40}}
\end{figure}

    Figures~\ref{fig37} to \ref{fig40} show the results of the simulations
along with some comparisons with the observations.  Figure~\ref{fig37} shows
the distribution of pixel intensities for the 140- and 240-$\mu m$ continuum
maps and for the $\cOone$ line map for both the simulations and observations.
The pixels represented in the histograms are only those where
$\Ia$, $\Ib$, and $\Ic$ are simultaneously greater than 5-$\sigma$.  This
corresponds to 1465 pixels for the simulations and 674 pixels for the 
observations.  Even after normalizing for the factor of $\sim$2 greater number 
of high signal-to-noise pixels in the simulations, the number of medium- and 
high-intensity pixels (i.e. $\gsim 200\MJsr$ for $\Ia$ and $\gsim 100\MJsr$ 
for $\Ib$) in the 140- and 240-$\mu m$ simulated maps is about 2 to 3 times
higher than for the maps of the real observations.  For the $\Ic$ map, the
simulations have about a factor of 5 higher number of pixels of medium- and 
high-intensity (i.e. $\Ic\gsim 2\Kkms$) than in the observations.   All the
simulated maps have a higher ratio of medium- and high-intensity pixels to 
low-intensity pixels than the observations.  This is especially true for the
$\cOone$ maps.  This is partly because the simulations have roughly twice the
fraction of low $\rd$ values than do the observations (i.e. for $\rd\lsim
20\MJkk$).  Nevertheless, the normalized pixel distributions of the simulations
agree with those of the observations to within factors of a few.  Exact 
agreement is not necessary in any case, because the purpose of the
simulations is to check how well the original input parameters are recovered,
whether those parameters adequately mimic the real observations or not.

    Another check of this mimicry is given in Figure~\ref{fig38}.  These are the 
plots of $\rd$ versus $\Tdc$ upon which all of the modeling in the current work 
is based.  The simulations adequately reproduce the main features of the 
observations: the triangular cluster of points for $\Tdc\lsim 21\,$K and the
monotonic rise for $\Tdc\gsim 20\,$K.  However, the simulations do {\it not\/}
account for the observed points that fill in the center of the triangular 
cluster and also do {\it not\/} account for the points of $\rd\gsim 80\MJkk$. 
This comparison between simulations and observations suggests that the basic 
assumption (see Paper~I) is not correct and that we need appropriately chosen 
subsamples, each with its own set of physical conditions, to account for the 
shortcomings in the simulations (see the end of Section~3.4 of Paper~I).  Nevertheless,
the simulated $\rd$ versus $\Tdc$ plot is an adequate representation of the
observations.  In fact, the noise in the simulations seems to account for the
low-$\rd$ points (i.e. the points with $\Tdc=18$ to 22$\,$K and $\rd\lsim
15\MJkk$) mentioned in Section~3.1 of Paper~I.

    Figure~\ref{fig39} further compares the simulations with
the observations, and has plots of the one-component, continuum-derived
gas column densities, $\NHd$, versus the dust temperature, $\Td$, in the 
one-component case.  Since these are continuum-derived quantities, they are 
independent of the particular parameter values of the one-component model 
(e.g., gas density, gas column density per velocity interval, etc.).  Again,
the simulations adequately imitate the observations.   There are only slight
differences.  For example, the simulations show a hook-like feature centered
at $\Td\simeq 17.5$, $\NHd\simeq 100$, which is nearly, but not completely, 
absent from the observations.  Another example is a spur that extends from
$\Td\simeq 14.3$ to 18 for $\NHd\simeq 15$ in the simulations that is only
hinted at in the observations.  Notice also that the simulations have a
smaller vertical spread in the $\Td>20\,$K points than do the observations.
Still, these are just minor discrepancies.

    Like Figure~\ref{fig39}, Figure~\ref{fig40} plots the continuum-derived
gas column densities against the dust temperature (the component-1 temperature
for this figure), but this time for the two-component, two-subsample models.  
For the two-component cases, the specific parameter values {\it do\/} indeed 
matter.  Specifically, the resultant parameter values from the model fits to the 
actual observations are those given in Table~2 of Paper~I. The resultant parameter 
values from model fits to the simulations are given in Table~\ref{tbl-7} (the model 
results from the data with noise).  Again, the simulations satisfactorily represent 
the observations and have only minor discrepancies.  The most noticeable of these is 
the group of points with large error bars at $\Tdo=3$ to 8 for $\NHd\simeq 150$ 
to 500 that occur for the real observations and are not in the simulations. 

    Given that the simulations are reasonable, we 
now examine how well the models recover the inputs.  We start with the
most realistic models --- the two-component, two-subsample, LVG models ---
and move towards the simplistic models --- the one-component models ---
to see what information they can realistically recover.

\subsection{Two-Component, Two-Subsample Models of the Simulations\label{sssec371}}

\begin{figure}
\epsscale{0.7}
\plotone{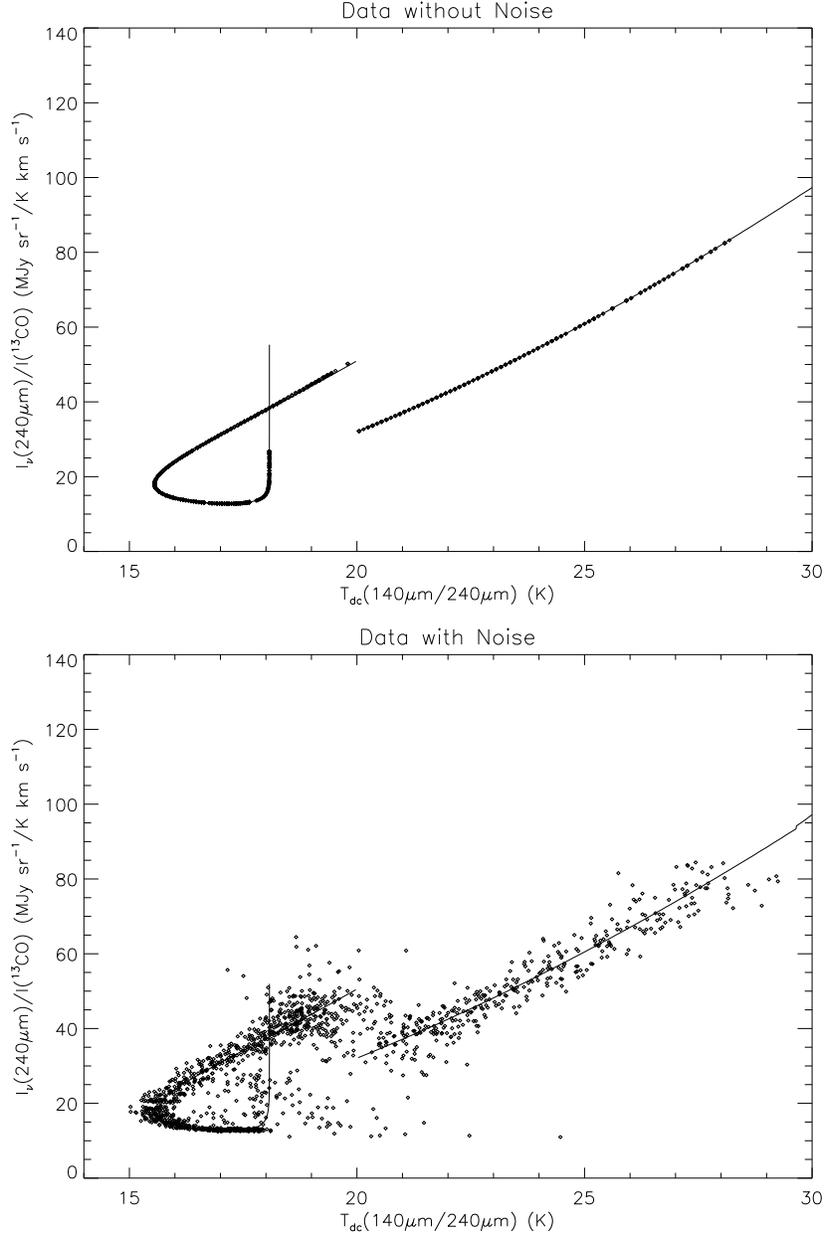}
\caption{Plots of $\rd$ versus the 140$\um$/240$\um$ color temperature are
shown for the simulations with{\it out} noise (upper panel) and with noise
(lower panel) along with the best-fit model curves for the two-component,
two-subsample models.  The parameter values used to generate these curves
are listed in Table~\ref{tbl-7}.   The panels only include those pixels with
intensities above the 5-$\sigma$ level in $\Ia$, $\Ib$, $\Ic$ {\it 
simultaneously\/}.  
\label{fig41}}
\end{figure}

    The best fitting model curves to the simulations for the two-component, 
two-subsample models are shown in Figure~\ref{fig41} and the corresponding
parameter values are given in the last four columns of Table~\ref{tbl-7}.
Columns~4 and 5 of Table~\ref{tbl-7} list the model results from fitting the
models to the data before the noise was added --- i.e., the noise-free data.
Columns~6 and 7 list those results for the fits to the data that have noise 
added.  The results in these columns can be compared with the simulation 
inputs in columns~2 and 3.  (Column~1 gives the parameter names.)  The two 
subsamples were chosen from those pixels for which the signal-to-noise ratio 
was $\geq 5$ in $\Ia$, $\Ib$, $\Ic$ simultaneously.  Of course, the 
signal-to-noise ratio is not defined for the noise-free maps; so the pixels 
that matched the signal-to-noise criteria in the maps with the added noise 
were also the pixels chosen in the corresponding noise-free maps.  Also, fitting
the model required specifying the error bars, even for the noise-free maps.  
The error bars were specified to be the same as those in the corresponding maps 
with added noise, even though the noise-free maps had no noise and, therefore, 
no errors. 

    A number of important conclusions result from comparing the
results with the inputs.  The most important is that {\bf completely} 
{\it recovering the inputs {\bf even} in the noise-free case is 
{\bf not} possible.\/}  This despite the model curves fitting the data extremely
well (see Figure~\ref{fig41}).  Accordingly, problems like not recovering 
the correct values of $c_0$, $n_{c0}$, or $n_{c1}$ within an order of magnitude 
or more are {\it intrinsic shortcomings\/} of the method itself and are {\it 
not\/} entirely due to the uncertainties caused by noise in the data. 
Also note that some results are {\it more\/} accurate in the 
noise-{\it added\/} data than in the noise-free data.  For example, $c_0$
for both the $\Tdc<20\,$K and $\Tdc\geq 20\,$K subsamples was more accurately
recovered in the model fits to the data with noise than in fits to the 
noise-free data.  This is also the case for $\nvcz$ for the $\Tdc<20\,$K
subsample.  Better recovery from fits to the data with noise is probably
just random luck.  As discussed in Section~3.3 of Paper~I, the fitting process
itself has random elements, such as the choice of starting grid.  This 
choice affects the final results of some parameters.  Consequently,
a different choice of starting grid could easily result in 
worse recovery than better.  

    Comparing the particular model results found here with the inputs gives 
a crude measure of the accuracy of the modeling.  The results of this
comparison are summarized below:
\begin{itemize}
\item $\DT$ is within 1$\,$K for the $\Tdc<20\,$K subsample and within
2$\,$K for the $\Tdc\geq 20\,$K subsample (within 1$\,$K in the noise-free
case). 
\item $\Tdz$ for the $\Tdc<20\,$K subsample is known within the formal 
uncertainty of $\leq 1\times 10^{-5}\,$K.  (For the actual observations 
this would be an order of magnitude larger.)   For the $\Tdc\geq 20\,$K 
subsample, the value for $\Tdz$ is adopted.
\item $c_0$ is known within a factor of 2 for the $\Tdc<20\,$K subsample
(within a factor of 16 for the noise-free case).  It is known within a
factor of 3 for the $\Tdc\geq 20\,$K subsample (within a factor of 2 for
the noise-free case).
\item $\nvcz$ is known within a factor of 3 for both subsamples (within
a factor of 10 for the noise-free case for the $\Tdc<20\,$K subsample
and within a factor of 2 for the noise-free case for the $\Tdc\geq 20\,$K 
subsample).
\item Again, the product $c_0\nvcz$ is more accurately recovered than either
of its factors.  This is known to within a factor of 2 for both subsamples
(also within a factor of 2 or exactly correct in the noise-free case 
depending on the subsample). 
\item $n_{c0}$ is out by 3 orders of magnitude or exactly correct
depending on the subsample (with the same behavior in the noise-free
case).
\item $\nvco$ is within a factor of 2 for both subsamples (within a
factor of 2 or exactly correct in the noise-free case depending on the
subsample).
\item $n_{c1}$ is within a factor of 6 for the $\Tdc<20\,$K subsample
and within a factor of 2 for the $\Tdc\geq 20\,$K subsample (within
a factor of 6 or exactly correct in the noise-free case depending on
the subsample).  
\end{itemize}
We discuss these accuracies in more detail after
examining the results of simple two-component models applied to the
simulated data in the next subsection.

    One important point is the
reliability of the $\DT$ result.  Given that the two-component model 
results {\it always\/} yield a $\DT$ value that is within 1$\,$K, or 
sometimes 2$\,$K, of zero, is it {\it possible\/} that the two-component
models {\it always\/} yield this result, {\it regardless of the true
value of $\DT$?\/}  This was tested by modeling simulated maps with
inputs $\DT=8\,$K for the $\Tdc<20\,$K subsample and $\DT=10\,K$ for the
$\Tdc\geq 20\,$K subsample. {\it The two-component, two-subsample model 
results were again within 1$\,$K of the input $\DT$ values.\/}  Therefore,
$\DT$ is very likely zero for the observations as well. 

\begin{figure}
\epsscale{0.7}
\plotone{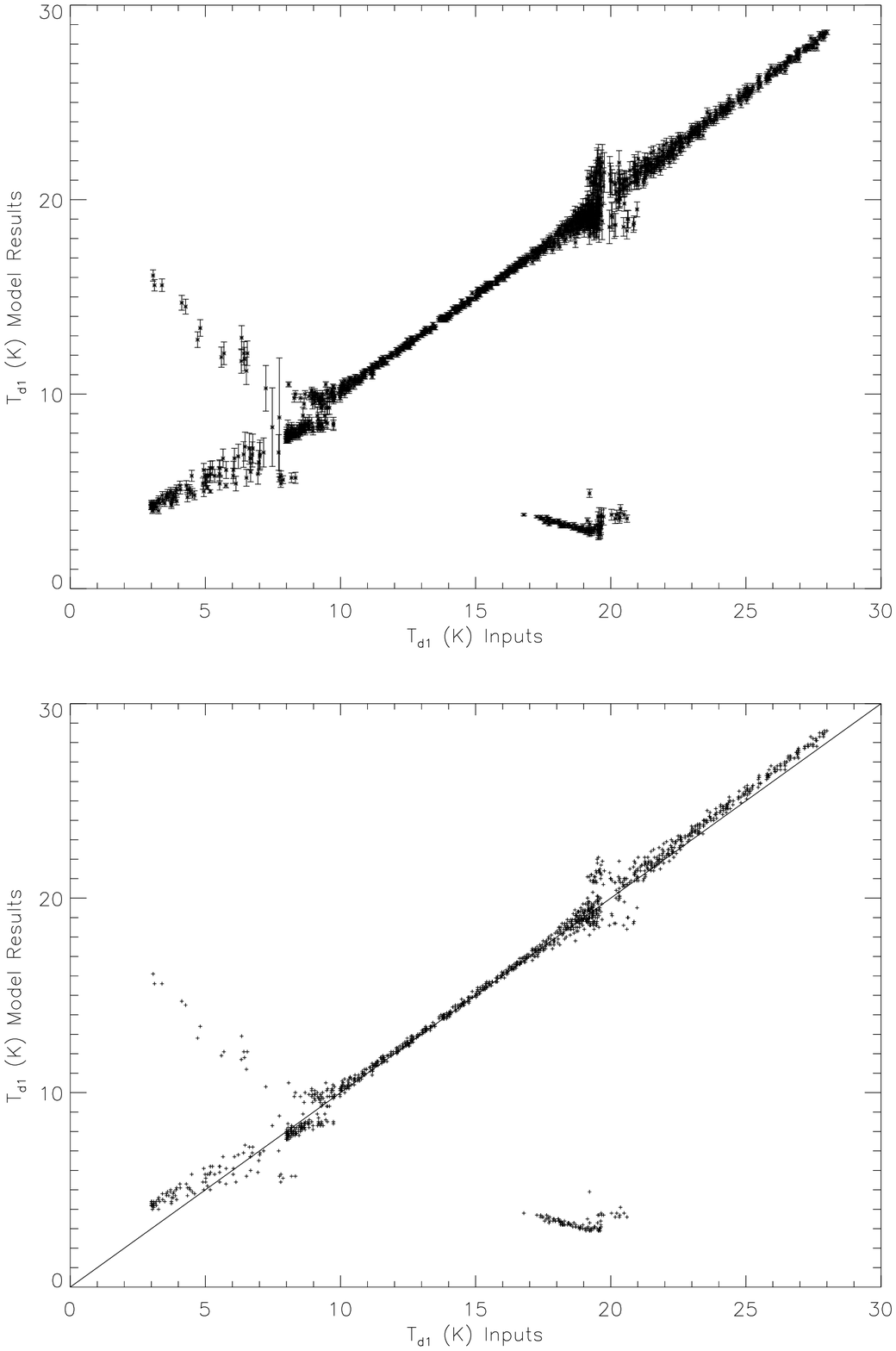}
\caption{The component-1 dust temperature as derived from fitting a 
two-component, two-subsample model is plotted against the component-1 dust 
temperature input values for the simulated data.  The upper panel includes 
the error bars in the model results, while the lower panel omits these error 
bars.  The lower panel also includes a solid straight line that represents
$\Tdo(model)=\Tdo(input)$ for comparison with the plotted points. 
The plots only include those pixels with the intensities above the 5-$\sigma$ 
level in $\Ia$, $\Ib$, $\Ic$ {\it simultaneously\/}. 
\label{fig42}}
\end{figure}

    The best fitting model curves can find the component-1 dust 
temperatures and the column densities as a function of position.  These are 
compared with the original input values.  Figure~\ref{fig42} shows 
the recovered $\Tdo$ values plotted against the input $\Tdo$ values.  Despite 
the noise in the simulated maps, the recovered $\Tdo$ values match the input 
values to within a few percent for the majority of (high signal-to-noise) 
points.  The most noticeable exceptions occur in two spurs that extend 
above and below the solid line plotted in the lower panel of that figure.  
The upper spur represents those positions where $\Tdo$ is between about 3 
and 8$\,$K, but has been misidentified as being between 16 and 9$\,$K.  The 
lower spur represents another misidentification of $\Tdo$, but in the opposite 
sense: $\Tdo$ is really between 17 and 20$\,$K, but has been assigned to be 
between 4 and 3$\,$K.  This mistake in assigning the correct $\Tdo$ value for 
some positions is easy to understand.  In Figure~\ref{fig41}, 
the model curve for the $\Tdc < 20\,$K sample crosses itself; there
is a vertical segment that crosses an inclined segment.  At the intersection 
point, the vertical segment has $\Tdo\simeq 3$-$4\,$K and the inclined 
segment has $\Tdo\simeq 18\,$K.  Therefore, any points in the $\rd$ versus
$\Tdc$ plot near this intersection point are easily misassigned to the
vertical segment, when it really belongs to the inclined segment, and
vice versa.  As the noise in the data grows larger, more points will
be assigned to the wrong segment.  In this case, the number of misassigned
points is only 8\% of the total number of high signal-to-noise points. 

\begin{figure}
\epsscale{0.65}
\plotone{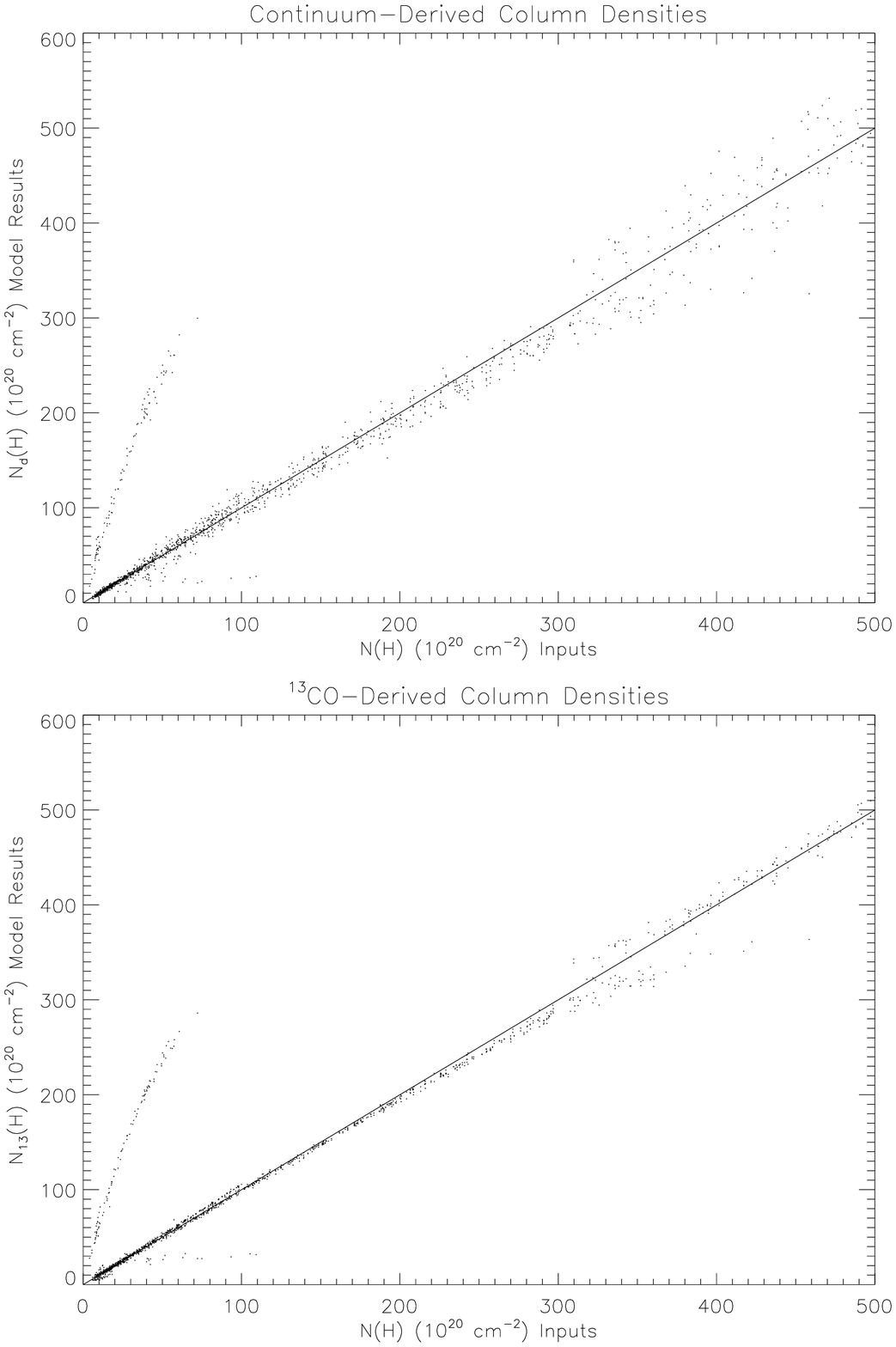}
\caption{Plots of the model gas column densities, $\NHd$ and $\Nnth$, for the 
two-component, two-subsample case versus the input column density values
are shown for the simulations.  All column densities are in units of $10^{20}\ 
H\ nuclei\cdot\unit cm^{-2}$.  The upper panel is the plot for the continuum-derived
column densities, $\NHd$, and the lower panel is for the $\cO$-derived 
column densities, $\Nnth$. The panels omit the error bars for clarity. The right 
panels also include a solid line representing the hypothetical case of agreement 
between the inputs and the model results (i.e. slope=1 and y-intercept=0). The 
panels only include those pixels with intensities above the 5-$\sigma$ level in 
$\Ia$, $\Ib$, $\Ic$ {\it simultaneously\/}.   
\label{fig43}}
\end{figure}

    The misassignment of $\Tdo$ values changes the determination
of column densities.  This is illustrated in the panels of Figure~\ref{fig43},
which are plots of the model-derived column densities (i.e., continuum-derived 
gas column densities and $\cO$ line-derived gas column densities) versus the 
input column densities.  As in the previous figure, the majority of positions 
show nearly perfect agreement (within a few percent) between the model-derived 
column densities and the input column densities.  However, again as in the 
previous figure, there are two spurs representing strong disagreements.  In
this figure the disagreements are factors of $\sim$4-6 in either direction.  
Obviously, the spurs in the column density plots of Figure~\ref{fig43}
correspond to the spurs in the dust temperature plots of Figure~\ref{fig42},
although in the opposite sense: the upper spur in the dust temperature plots
corresponds to the lower spur in the column density plots and vice versa.  
The question is why the disagreements are around a factor of 5.  Starting with 
equation~(40) of Paper~I,  we first consider the
case where the $\Tdo$ of a position is 18$\,$K, which is numerically equal
to $\Tdz$, and has been misassigned to 4$\,$K.  If the $\Tdo$ value had 
been correct, then the correct column density {\it would\/} have been given 
by
\begin{equation}
\NHd(correct) ={f_{\nu_{10}}(\Tdc)\over f_{\nu_{10}}(\Tdz)}\ \NHdo\qquad ,
\label{mr46}
\end{equation}
which was obtained by setting $\Tdo=\Tdz$.  This in turn implies $\Tdc=\Tdz$
and (\ref{mr46}) simplifies to
\begin{equation}
\NHd(correct) = \NHdo\qquad .
\label{mr47}
\end{equation}
But, because this data point has $\Tdo$ misassigned to some low value, we
have
\begin{equation}
\NHd(incorrect)\simeq {\left[\nvtco\right]_{\hbox{$_{mod}$}} + 
c_0 \left[\nvtcz\right]_{\hbox{$_{mod}$}}\over 
c_0\left[\nvtcz\right]_{\hbox{$_{mod}$}}}\ \,
{f_{\nu_{10}}(\Tdc)\over f_{\nu_{10}}(\Tdz)}\ \,\NHdo\qquad ,
\label{mr48}
\end{equation}
where $f_{\nu_{10}}(\Tdo) << f_{\nu_{10}}(\Tdz)$ was assumed.  This 
assumption is especially valid in the Wien limit, which applies to
the 240$\um$ continuum for these temperatures.  Because the Wien limit
applies, we can also state that $\Tdc\simeq\Tdz$, so that 
$f_{\nu_{10}}(\Tdc)\simeq f_{\nu_{10}}(\Tdz)$.  Using this and dividing
(\ref{mr48}) by (\ref{mr47}) yields
\begin{equation}
{\NHd(incorrect)\over\NHd(correct)}\simeq 
{\left[\nvtco\right]_{\hbox{$_{mod}$}} + 
c_0 \left[\nvtcz\right]_{\hbox{$_{mod}$}}\over 
c_0\left[\nvtcz\right]_{\hbox{$_{mod}$}}}\qquad .
\label{mr49}
\end{equation}
Using the parameter values in Table~\ref{tbl-7} for the $\Tdc<20\,$K subsample
for the data with noise gives $\NHd(incorrect)/\NHd(correct)\simeq 5$ as
desired.  For the real observed data, the model parameter values in
Table~2 again give $\NHd(incorrect)/\NHd(correct)\simeq 5$. 
Note that in the opposite case where the $\Tdo=4\,$K data point is misassigned
to $\Tdo=18\,$K, the right side of expression~(\ref{mr49}) is changed to 
its reciprocal or, numerically, 0.2.  Even though there are two spurs, there
are many more points in the upper spur than in the lower spur; this results
in overestimate of the total mass of about 7\%. 

    The above only explains the spur locations in the continuum-derived 
column density plots of Figure~\ref{fig43} (i.e., the upper panels).  The 
explanation for the $\cO$-derived column densities is similar.  Instead of 
starting with expression~(40) of Paper~I as was done for the continuum-derived 
column densities, we would start with (34) of Paper~I.  Since $\Tr\propto\NDv$ for 
a large area of parameter space (see Section~3.3 of Paper~I), the arguments used 
above apply to the $\cO$-derived column densities as well.  The only difference 
is that, now, the Wien limit does not apply and we may not be able to approximate 
the $(\Trone+$$c_0\Trz)$ in the denominator with $c_0\Trz$.
Nevertheless, such an approximation is {\it still} valid, because these
radiation temperatures are with respect to the cosmic background temperature
of roughly 3$\,$K.  So, for the example discussed here, where $\Tdo=18\,$K
is mistaken for $\Tdo=4\,$K, $\Trone/\Trz$ is not ${4\over 18}$, but 
closer to ${1\over 15}$, more than 3 times smaller.  Consequently, 
equation~(\ref{mr49}) and its reciprocal are still valid for the $\cO$-derived
column densities.  The overestimate of the total mass from using the $\cO$
data is similar to that for the continuum-derived total mass: 6\%.

\begin{figure}
\epsscale{0.67}
\plotone{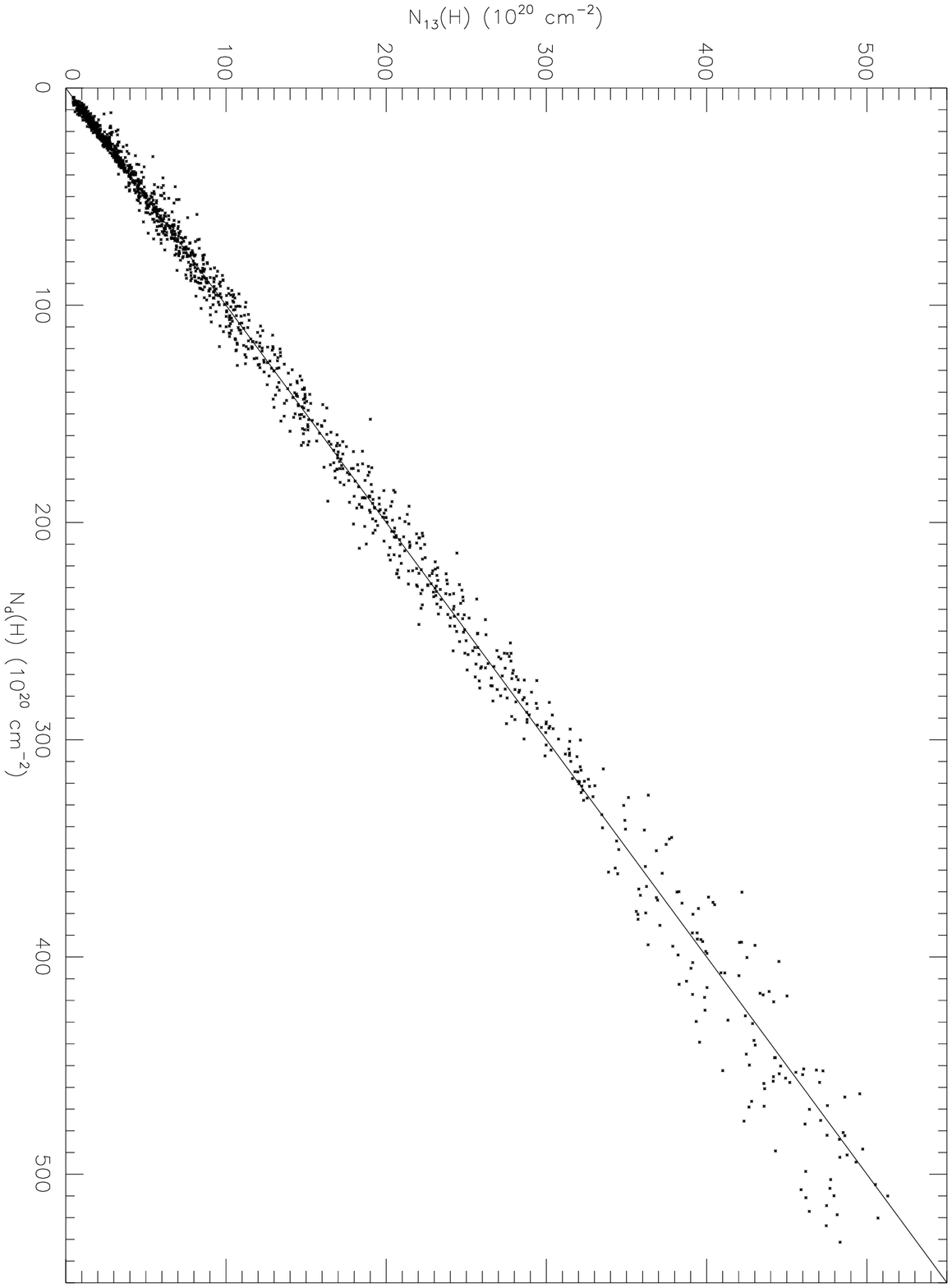}
\caption{Plot of the continuum-derived gas column densities, $\NHd$, versus
the $\cO$-derived gas column densities, $\Nnth$, is shown for the simulations,
where the column densities were derived using the parameters from the best-fit
two-component, two-subsample models. 
All column densities are in units of $10^{20}\ H\ nuclei\cdot\unit cm^{-2}$.    
The plot includes a solid straight line that 
represents $\Nnth=\NHd$ for comparison with the plotted points. 
The plots only include those pixels with the intensities above the 5-$\sigma$ 
level in $\Ia$, $\Ib$, $\Ic$ {\it simultaneously\/}.  The error bars are
omitted for clarity.
\label{fig44}}
\end{figure}

    Figure~\ref{fig44} shows that the two types of model-derived column 
densities agree with each other extremely well, {\it despite\/} having
7\% of these wrong by factors of 5.  The total masses also agree well 
because the erroneous column densities are wrong by the same factors for 
both the continuum-derived and $\cO$-derived column densities. 

    In summary, the simulations show that even modeling the noise-free data
will not allow perfect recovery of the parameters.  Nevertheless, the simulations
show that we obtain $\DT$ to within 1 or 2$\,$K (even when that $\DT$ is different
from zero), $\Tdz$ to better than a millikelvin for the $\Tdc < 20\,$K subsample, 
the component-0 density can be off by 3 orders of magnitude, and the other parameters 
might be known to within about an order of magnitude.  Recovery of other quantities 
like the component-1 dust temperatures and the gas column densities is apparently 
accurate to within a few percent for 93\% of the points.  The other 7\% of the points 
have column densities too high or too low by a factor of about 5.  This results in 
overestimate of 6-7\% in the total mass.

\subsection{Simple Two-Component Models of the Simulations\label{sssec372}}

\begin{figure}
\epsscale{0.7}
\plotone{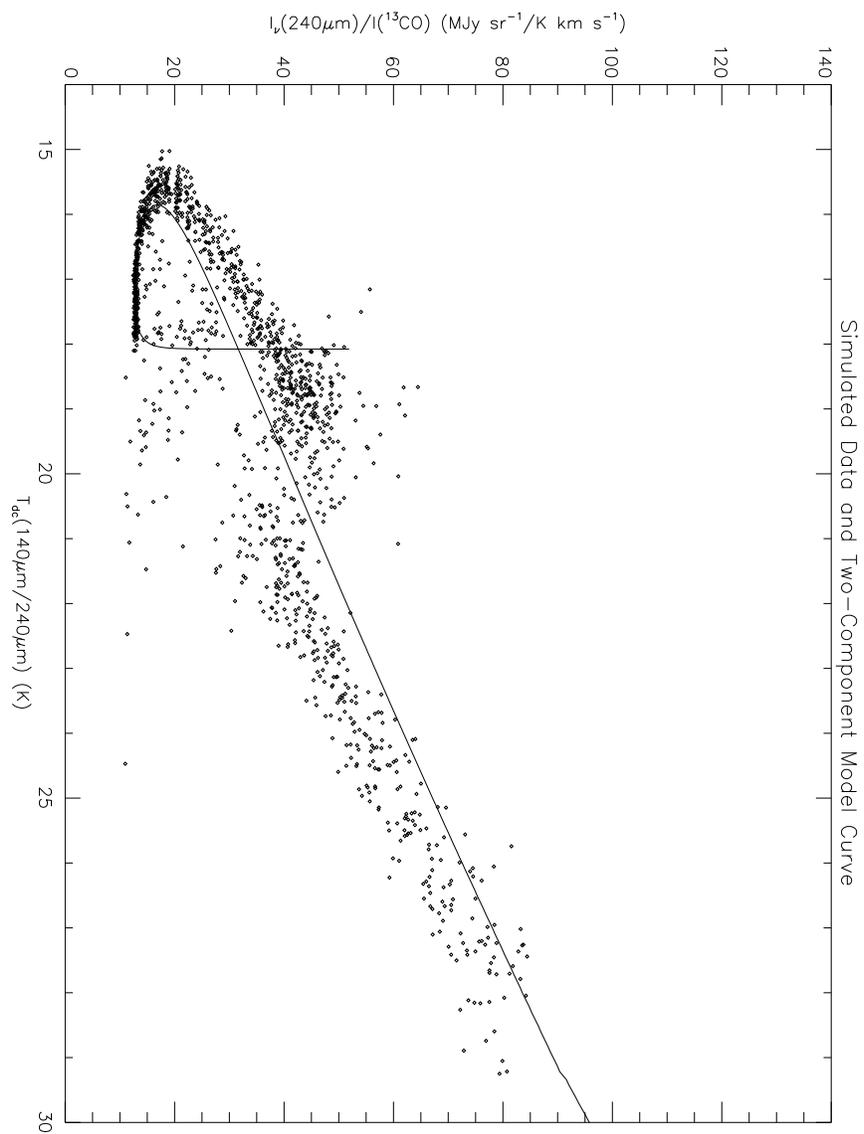}
\caption{Plot of $\rd$ versus the 140$\um$/240$\um$ color temperature is
shown for the simulations along with the best-fit model curves for the 
two-component models.   The plots only include those pixels with the intensities 
above the 5-$\sigma$ level in $\Ia$, $\Ib$, $\Ic$ {\it simultaneously\/}.  Error 
bars have been omitted for clarity. 
\label{fig45}}
\end{figure}

\begin{figure}
\epsscale{0.75}
\plotone{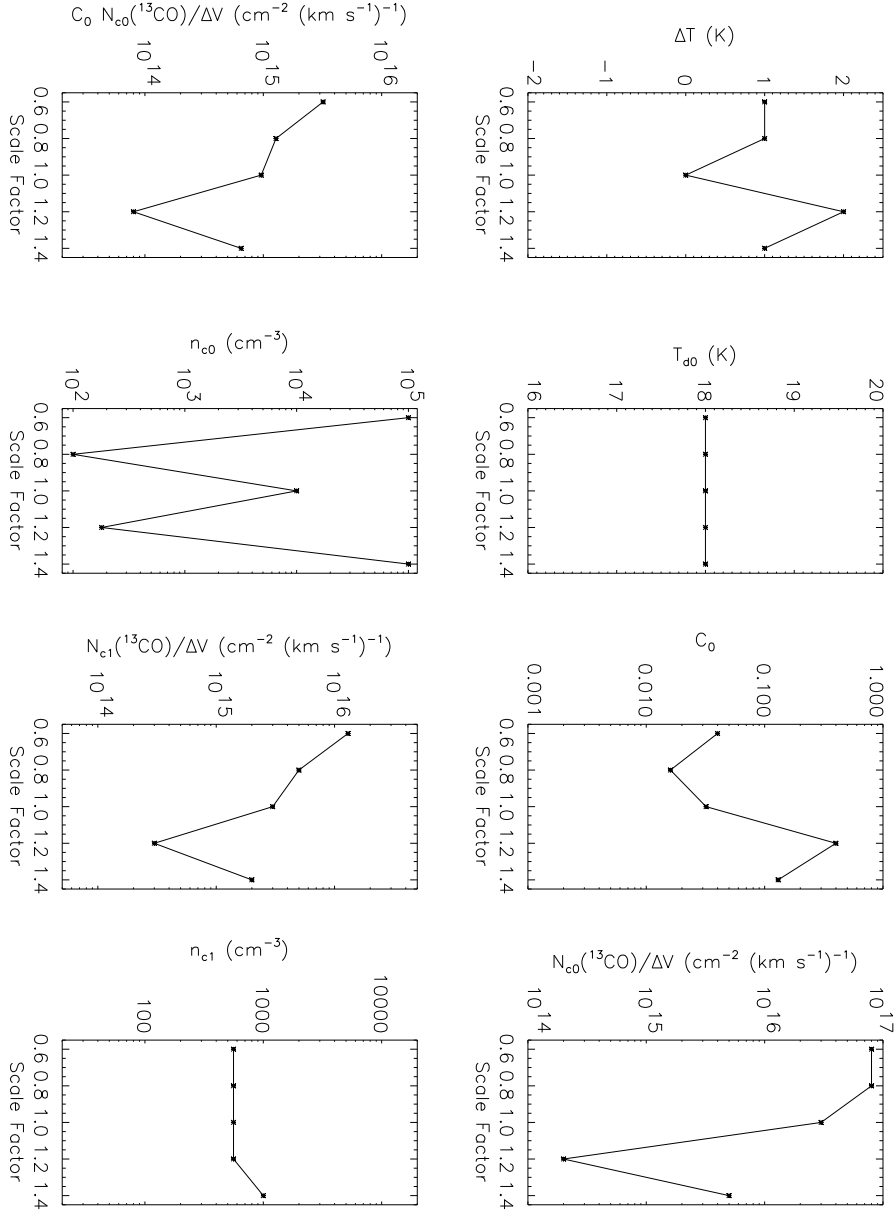}
\caption{The effect of the systematic uncertainties on the resultant
parameters from the fits of the two-component, LVG model curves to the
simulated data is shown.   The effect of these uncertainties was tested 
by applying the scale factors 0.6, 0.8, 1.0, 1.2, and 1.4 to the model 
curves and fitting the parameters for each scale factor.  Except for the 
plots for $\DT$ and $\Tdz$, all plots are semi-logarithmic where the 
vertical axes cover the about the same logarithmic difference in range 
(about 3 orders of magnitude).  This allows easy visual determination of 
which parameters have the smallest systematic uncertainties.  
\label{fig46}}
\end{figure}

    The best fitting model curve for the two-component models applied to
the whole sample of high signal-to-noise points in the simulations is shown 
in Figure~\ref{fig45}.  Again, these points corresponded
to those pixels for which the signal-to-noise ratio was $\geq 5$ in $\Ia$, 
$\Ib$, $\Ic$ simultaneously.  As done in Section~3.3 and Figure~21 of Paper~I,
Figure~\ref{fig46} shows the systematic effects on the resultant parameters
when a scale factor applied to the data is changed.  Comparing the various
panels of Figure~\ref{fig46} with the corresponding panels of Figure~21 of Paper~I
reveals strong similarities between the models applied to the 
simulations and those applied to the observed data.  The range of parameter 
variations is nearly identical in the two cases.  However, there is one
important difference between the systematic effects on the simulated data model 
results and those of the observed:  with the 
simulated data we can specify the accuracy of the
recovered results by comparing the ``true" values (i.e., the
inputs) with the model results;  with the actual observed data we can only 
estimate such accuracy by comparing the results in different 
cases (i.e., with different scale factors applied to the data or with different 
starting grids) with each other.  The accuracy of the recovered results for the 
simulations can also be tested by comparing the results in different cases
--- as was done in Figure~\ref{fig46}.  By comparing this accuracy with the 
accuracy obtained from comparisons with the input values, we now 
have insights into the estimated accuracies of the actual observations. 

    An example of such comparisons is inspecting how $\DT$ varies in 
Figure~\ref{fig46} about the $\DT$ value for a scale factor of unity (i.e. 
SF=1.0) and then comparing this with how those $\DT$ values vary about the original 
input value.  This then tells us whether the variation of $\DT$ with the scale 
factor for the real observations (see Figure~21 of Paper~I) 
is a realistic measure of the uncertainty in $\DT$.  In Figure~\ref{fig46},
$\DT$ varies within 2$\,$K of the value, i.e. $\DT=0\,$K, for SF=1.0.  The 
input value was $\DT=0\,$K.  Therefore, the variation of $\DT$ with the scale 
factor provides a reasonable estimate of the actual uncertainty in $\DT$.  For the 
models applied to the observations, Figure~21 of Paper~I shows us that $\DT$ varies 
within 1$\,$K of the value corresponding to SF=1.0, i.e. $\DT=0\,$K.  So we can say 
that the model $\DT$ value is within 1 or 2$\,$K of the true $\DT$ value.  Using the 
same arguments applied to $\Tdz$ suggests that this is known to within a 1$\,$mK
or less; this is undoubtedly optimistic and is dependent on the basic assumption.  
(It is also dependent on other assumptions, such as whether the spectral emissivity 
index, $\beta$, really is 2.0 or something nearby.  Paper~III suggests that
$\Tdz$ can be anywhere from $\sim$16 to $\sim$19$\,$K.)  For the other parameters, 
which had different input values for the two subsamples, we will compare the model 
results with the geometric mean of the two inputs:
\begin{itemize}
\item  The parameter $c_0$ can be off by a factor of 10 from the value corresponding 
to SF=1.0 (for both the simulations and the observations), but is off by a factor of 
40 from the input.  In addition, the model-derived $c_0$ values for all the scale 
factors are systematically lower than the input.  In other words, we cannot rely on 
varying the scale factor to give us parameter values that will surround the true 
value.  Again, $c_0$ is much more reliable when combined with $\nvcz$.  
\item $\nvcz$ itself can be out by a factor of 100 from the value corresponding to 
SF=1.0 (for both simulations and observations), {\it and\/} is {\it also\/} wrong by 
this factor compared to the input.  
\item The product $c_0\nvcz$ is off by at most only a factor of 10 compared to the 
value corresponding to SF=1.0, {\it and\/} is {\it also\/} wrong by this factor 
compared to the input.  Also, unlike $c_0$ alone, the range of different values of 
$c_0\nvcz$ corresponding to different scale factors does indeed include the input 
value. 
\item $n_{c0}$ is as much as a factor of 100 away from the value for SF=1.0 for the
simulations, and as much as factor 1000 away for the observations.  The different
$n_{c0}$ values for the simulations can be wrong by as much as a factor of 200 from 
the input, and the range of these values includes the input value.   It would seem, 
then, that the observations would suggest a greater uncertainty in $n_{c0}$ than 
would the simulations.
\item $\nvco$ is as much as a factor of 10 from the value at SF=1.0 for both the 
simulations and observations.  $\nvco$ can be as far as a factor of 30 from the
input, which is worse than the comparison with the $\nvco$ value at SF=1.0 would
suggest.  The range of possible $\nvco$ values for the different scale factors 
(see Figure~\ref{fig46}) includes the input value.
\item $n_{c1}$ can be as far as a factors of 2 or 3 from the value at SF=1.0 for
both the simulations and observations.  However, $n_{c1}$ can be out by a factor 
of 20 from the input value, much worse than comparison with the value at SF=1.0 
implies.  Also, another problem is that the range of possible $n_{c1}$ values 
(see Figure~\ref{fig46}) does {\it not\/} include the input: the model-derived
densities are all systematically too low by more than an order of magnitude.
\end{itemize}
The most interesting conclusion is that some parameters like
$c_0$ and $n_{c1}$ have a range of values that does {\it not\/} include the true
input value.  As mentioned previously, $c_0$ is assessed more reliably as
part of the $c_0\nvcz$ product, whose range of values does indeed include the
input value.  $n_{c1}$ still has this disadvantage, which cannot be ``fixed''
as easily as for $c_0$. 

    Based on the comparisons of the different results, the ranges of likely 
values of the different parameters have been listed in Table~\ref{tbl-8}. 
The range of values for each parameter assumes the minimum and
maximum values as in the case of the simple, and the two-subsample, 
two-component models --- with some important exceptions.  In 
the case of $\DT$, the maximum
value found was $+1\,$K, but the simulations suggest that $+2\,$K is also
possible.  Therefore, $+2\,$K is listed.   Note also that even though, for
simplicity, a one-component model was applied to the $\Tdc\geq 20\,$K subsample,
the two-component model results for that subsample represent the 
likely ranges listed in Table~\ref{tbl-8}.  For $c_0$ and $\nvcz$,
only the range of their product was listed, in order to provide more realistic
constraints on these parameters.  For the column density per velocity interval
in general, it was stated in Section~3.2 of Paper~I that the lower limit had 
to be about $3\times 10^{15}\,\cOit\, molecules\ckms$ as roughly constrained by the 
large-scale properties of the cloud.  For the two-component models, this lower
limit would apply to $\nvtco + $$c_0\nvtcz$.  However, $\nvtco$ is larger than
$c_0\nvtcz$ by factors of 3 to 4.  Therefore, the first term in that expression
dominates and it is sufficient to apply that lower limit to $\nvtco$ only, as
was done in Table~\ref{tbl-8}.  As for the densities, $n_{c0}$ and $n_{c1}$,
putting upper limits on those is not possible, because the results
are not distinguishable from those of LTE. 
Consequently, only lower limits are used.  Also the lower limit of $n_{c1}$
has been increased by an order of magnitude, because, as stated in the
previous paragraph, all the values of $n_{c1}$ found by the simple two-component
models are too low by at least an order of magnitude. 
 
\begin{figure}
\epsscale{0.7}
\plotone{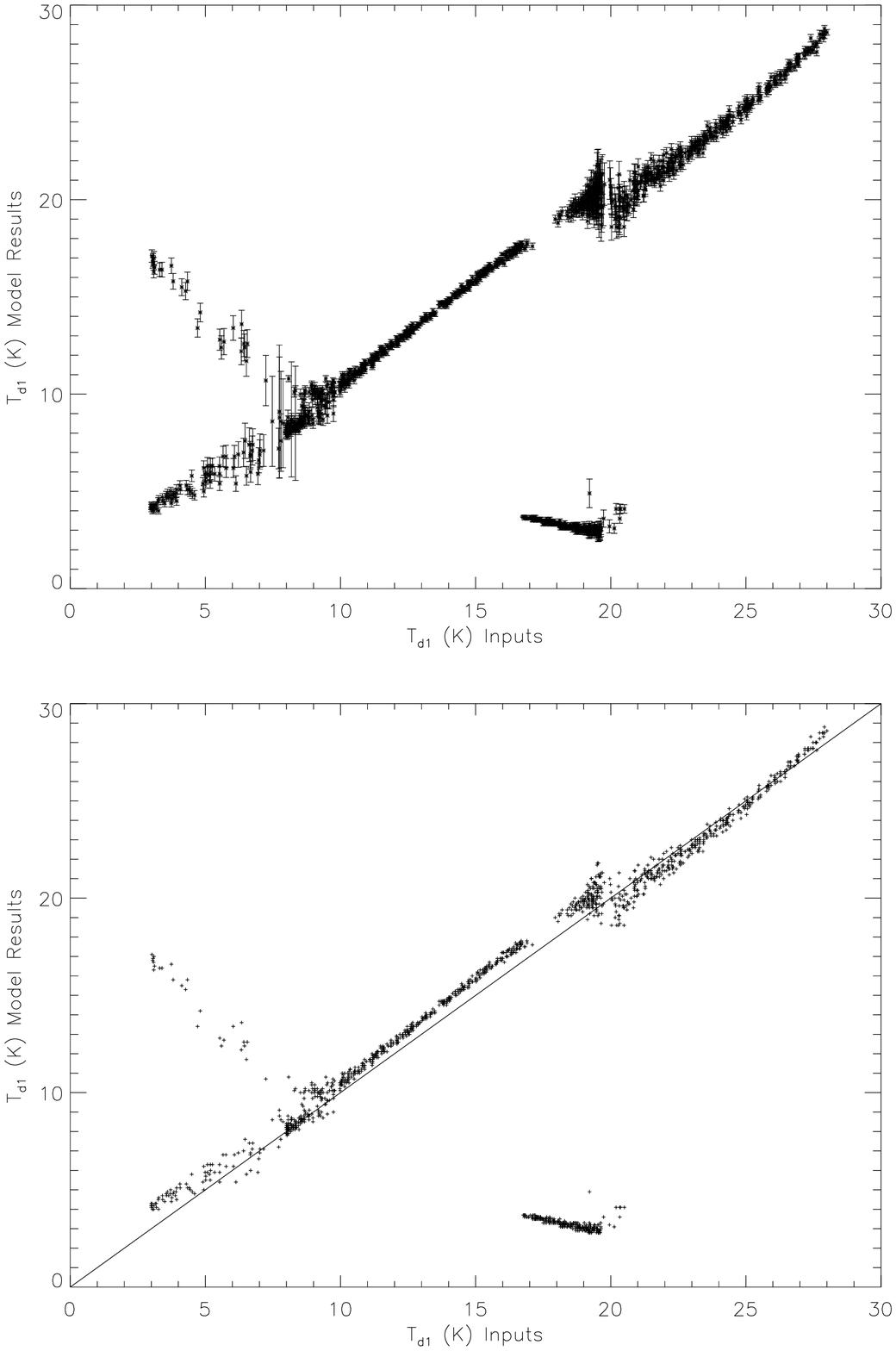}
\caption{The component-1 dust temperature as derived from fitting a 
two-component model is plotted against the component-1 dust temperature
input values for the simulated data.  The upper panel includes the 
error bars in the model results, while the lower panel omits these error 
bars.  The lower panel also includes a solid straight line that represents
$\Tdo(model)=\Tdo(input)$ for comparison with the plotted points. 
The plots only include those pixels with the intensities 
above the 5-$\sigma$ level in $\Ia$, $\Ib$, $\Ic$ {\it simultaneously\/}.
\label{fig47}}
\end{figure}

    The best fitting model curves shown in Figure~\ref{fig45} were used 
to find the component-1 dust temperatures and the column densities
as a function of position.  These are compared with the original
input values.  Figure~\ref{fig47} shows the recovered $\Tdo$ values plotted
against the input $\Tdo$ values.  Again, as in Figure~\ref{fig42}, the 
majority of recover $\Tdo$ values match the input values reasonably well,
except for the two spurs.  The noticeable difference, however, is the
systematic overestimate of $\Tdo$ for input $\Tdo$ values $\lsim 20\,$K
and a systematic underestimate of most of the $\Tdo$ values above this limit.
These systematic effects are obviously the result of forcing a single curve
to fit the two different subsamples: the curve systematically underestimates
a large fraction of the $\Tdc < 20\,$K subsample and overestimates most of
the $\Tdc\geq 20\,$K subsample. This results in systematically underestimating
(overestimating) the $\Tdo$ values for the simulated data points in the 
$\Tdo < 20\,$K ($\Tdc\geq 20\,$K) subsample.

\begin{figure}
\epsscale{0.65}
\plotone{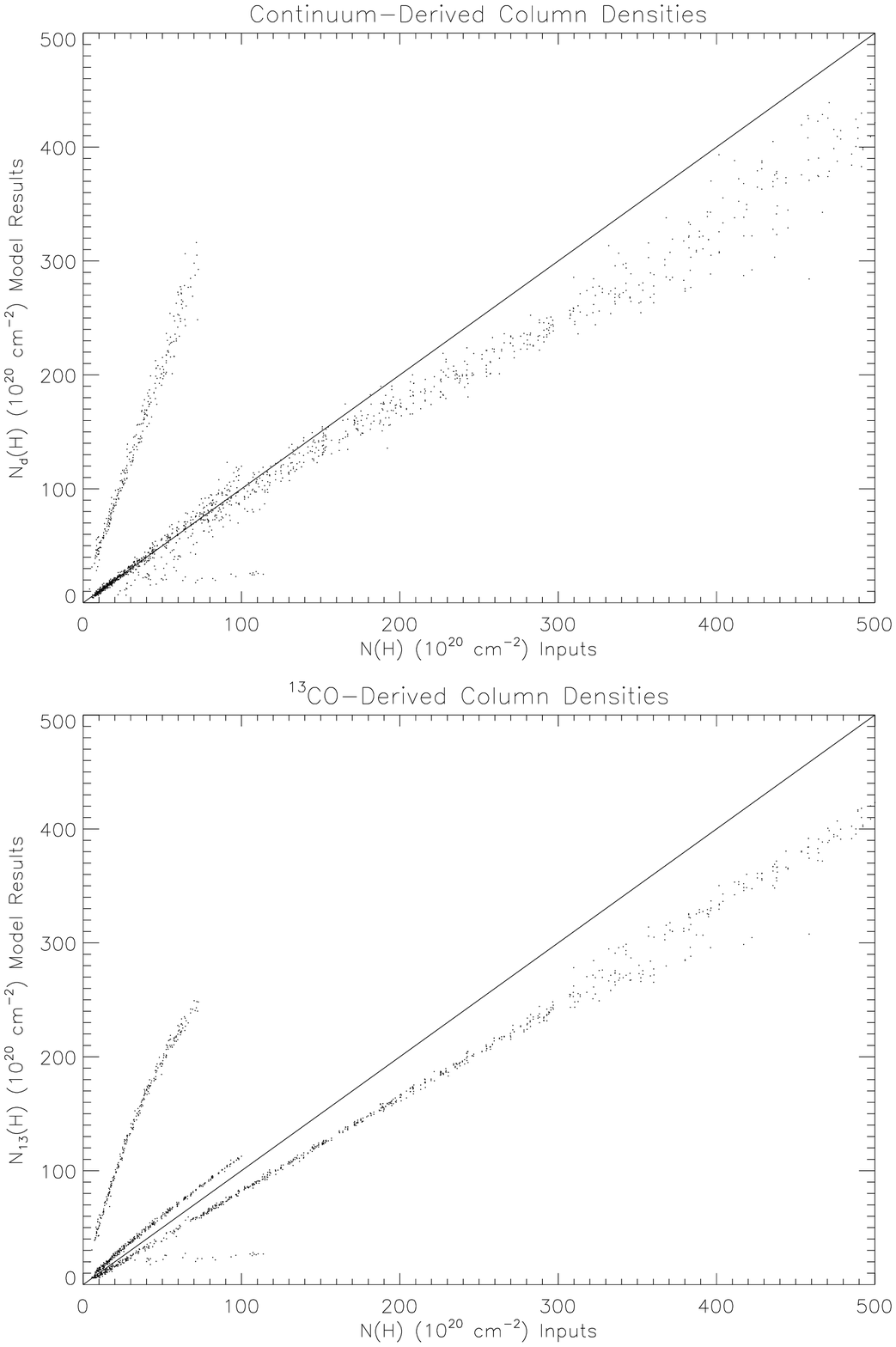}
\caption{Plots of the model gas column densities, $\NHd$ and $\Nnth$, for the 
two-component case versus the input column density values are shown for the 
simulations.  All column densities are in units of $10^{20}\ H\ nuclei\cdot\unit cm^{-2}$.  
The upper panel is the plot for the continuum-derived column densities, 
$\NHd$, and the lower panel is for the $\cO$-derived column densities, 
$\Nnth$. The panels omit the error bars for clarity.   A solid line is included
in each panel that represents the hypothetical case of agreement between the inputs 
and the model results (i.e. slope=1 and y-intercept=0).  The panels only include those 
pixels with intensities above the 5-$\sigma$ level in $\Ia$, $\Ib$, $\Ic$ {\it 
simultaneously\/}.   
\label{fig48}}
\end{figure}

\begin{figure}
\epsscale{0.67}
\plotone{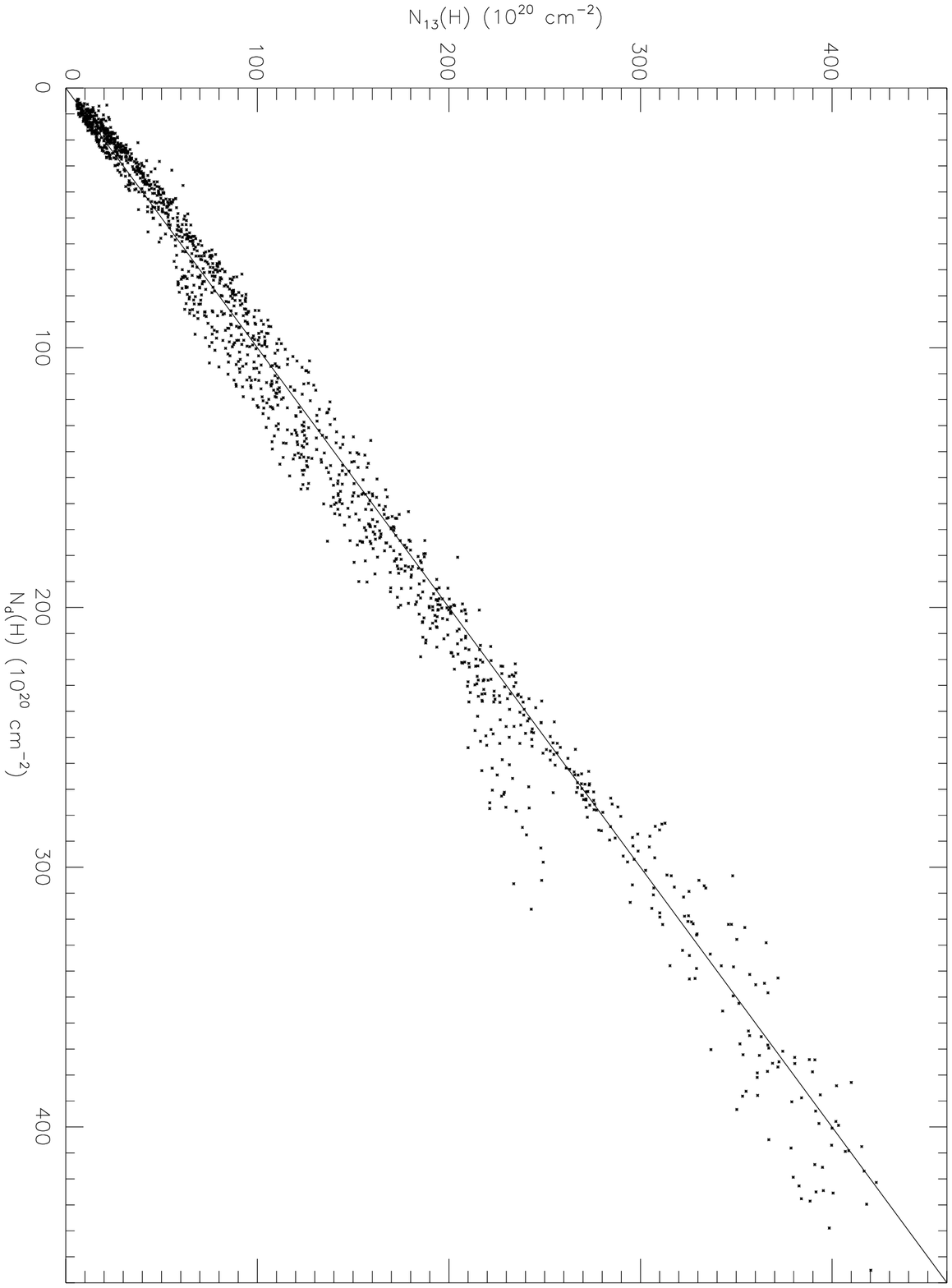}
\caption{Plot of the continuum-derived gas column densities, $\NHd$, versus
the $\cO$-derived gas column densities, $\Nnth$, is shown for the simulations,
where the column densities were derived using the parameters from the best-fit
two-component models.  All column densities are in units of $10^{20}\ H\ nuclei 
\cdot\unit cm^{-2}$.   A solid straight line is included that represents 
$\Nnth=\NHd$ for comparison with the plotted points.  The plots only include 
those pixels with the intensities above the 5-$\sigma$ level in $\Ia$, $\Ib$, 
$\Ic$ {\it simultaneously\/}.  Error bars are omitted for clarity. 
\label{fig49}}
\end{figure}

    The incorrect estimates of the $\Tdo$ values change the 
determination of the column densities.  This is obvious in the panels of 
Figure~\ref{fig48},  which are plots of the model-derived column densities 
versus the input column 
densities, analogous to those in Figure~\ref{fig43} for the two-component,
two-subsample models.  As in Figure~\ref{fig43}, there are two spurs of
very large disagreements (i.e., factors of $\sim 5$ in both directions).
But, unlike that figure, Figure~\ref{fig48} shows systematic disagreements
of about 10\% and 20\% on either side of the solid line --- the line that
represents perfect agreement.  Again those disagreements follow naturally
from the disagreements seen in the plot of $\Tdo$ values in Figure~\ref{fig47}:
the points that have overestimated $\Tdo$ values in the $\Tdc < 20\,$K 
subsample will have underestimated column densities and vice versa for many
of the points in the $\Tdc\geq 20\,$K.  Despite these noticeable disagreements,
the continuum-derived and $\cO$-derived column densities in
Figure~\ref{fig49} agree well, although with noticeably larger
scatter than in Figure~\ref{fig44} for the two-component, two-subsample
models.  Also, the total mass estimated from the model results is only
overestimated by about 3 to 6\%. 

    In summary, the results of the simple two-component models applied to the 
simulations for different scale factors has allowed reasonable estimates of the 
ranges of parameter values for all the two-component models.  These ranges allow for 
systematic uncertainties in the real observations and are listed in Table~\ref{tbl-8}.
There are noticeable systematic errors in the derived component-1 dust temperatures
and in the derived column densities.  Despite these systematic errors, the simple
two-component models still give reasonable estimates of the total
mass of the Orion clouds.

\subsection{One-Component, Non-LTE Models of the Simulations\label{sssec373}}

\begin{figure}
\epsscale{0.7}
\plotone{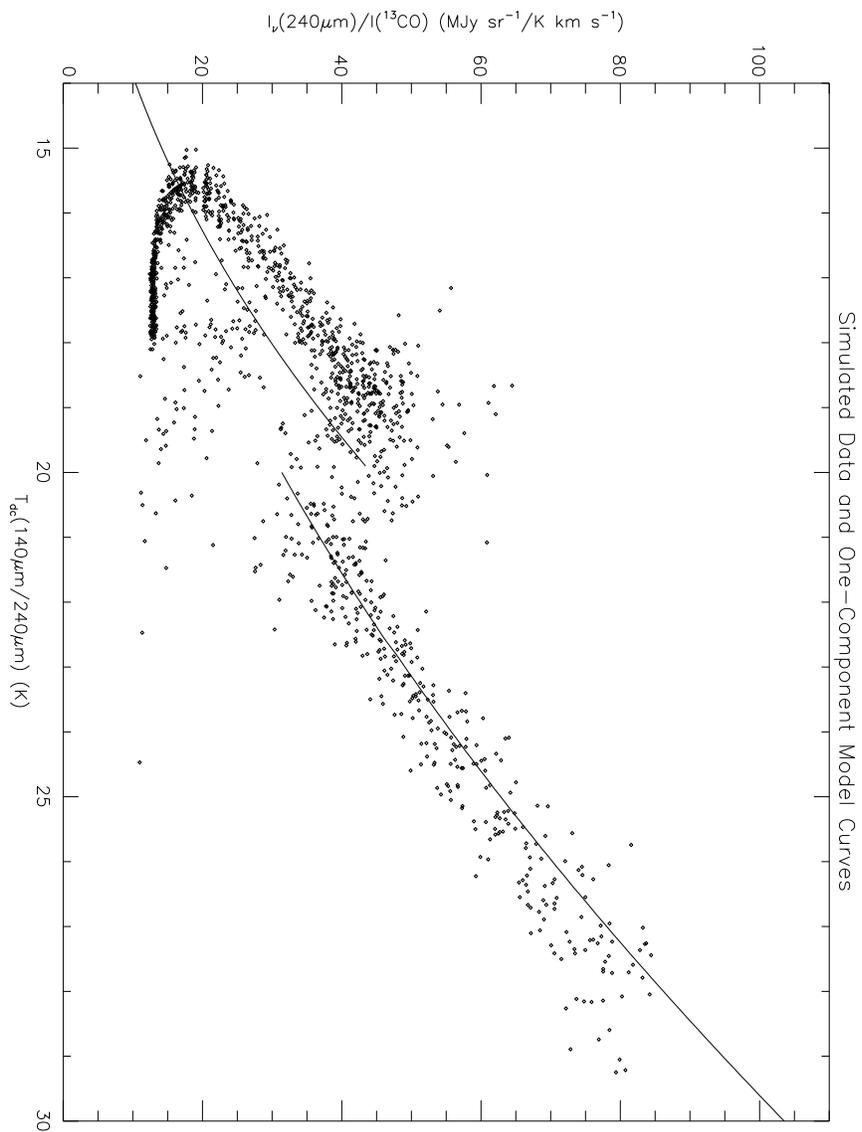}
\caption{Plot of $\rd$ versus the dust temperature is shown for the simulations 
along with the best-fit model curves for the one-component models.  These models
were applied to the $\Td<20\,$K and $\Td\geq 20\,$K subsamples separately.  The 
plots only include those pixels with the intensities above the 5-$\sigma$ level 
in $\Ia$, $\Ib$, $\Ic$ {\it simultaneously\/}.  Error bars have been omitted for 
clarity.  
\label{fig50}}
\end{figure}

\begin{figure}
\epsscale{0.62}
\plotone{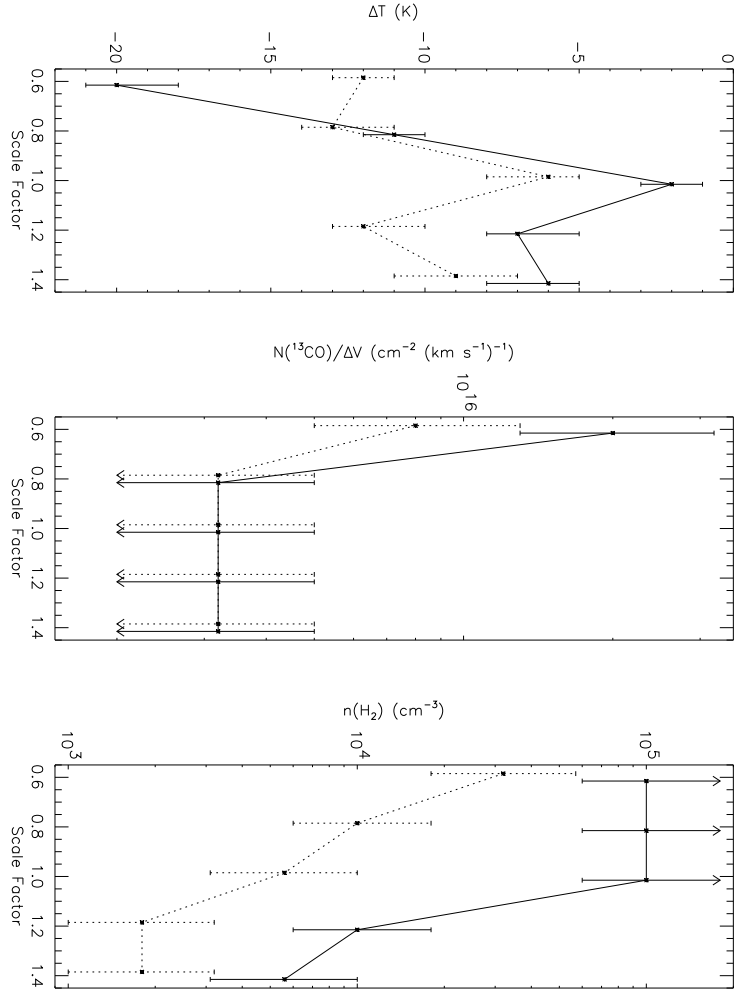}
\caption{The effect of the systematic uncertainties on the resultant
parameters from the fits of the LVG model curves to the simulated data is 
shown.  The effect of these uncertainties was tested by applying scale 
factors to the model curves and fitting the parameters for each scale factor.  
The left panel shows the resultant $\DT$ values, the center panel
shows the resultant $\NDv$ values, and the right panel shows the $\nH$ values.
The solid line in each panel represents the resultant parameter values for the 
fits to the subsample of points with $\Td<20\,$K.  The dotted line represents 
the resultant parameter values for the fits to the subsample of data with 
$\Td\geq 20\,K$.  Notice that the plotted points have been slightly 
displaced horizontally from their true scale factor values for clarity. The 
error bars represent the formal error bars for each model fit and are the 
minimum grid spacing, for the grid of LVG models used, necessary to increase 
$\chi^2$ by a {\it minimum\/} of $\chi_\nu^2$.  These formal errors are 
therefore very conservative estimates of the true formal errors.   
\label{fig51}}
\end{figure}

    The best fitting model curve for the one-component models applied to
the high signal-to-noise points in the simulations is depicted in
Figure~\ref{fig50}.  As discussed in Section~3.2 of Paper~I and illustrated in 
Figure~16 of Paper~I, Figure~\ref{fig51} of the current paper shows the systematic 
effects on the resultant parameters when a scale factor applied to the data is 
changed.  Comparing the three panels of Figure~\ref{fig51} with the corresponding 
panels of Figure~16 of Paper~I reveals that the models of 
the simulations and those of the observed data are similar.  The range of variation
of the parameters is nearly identical in the two cases, except in the panels
of the $\DT$ values: in that panel the model results of the simulations
have systematically lower $\DT$ values than those of the observations by 1 to
5$\,$K, the larger difference applying to the $\Td\geq 20\,$K subsample.  For
this subsample, the observed data points have lower $\rd$ values on average
than do the simulated data points.  

Also, the one-component modeling of the observed data was done a little 
differently from that of the simulated data.  The observed data were modeled
with the one-component models applied to the whole sample of points and
then again for just the $\Td\geq 20\,$K points.  In contrast, the simulated
data were modeled with the one-component models applied to {\it just\/} the
$\Td<20\,$K points and then {\it just\/} the $\Td\geq 20\,$K points.  In
short, the $\Td < 20\,$K subsample was {\it not\/} treated separately for
the observed data points, but was indeed treated separately for the simulated
data points. This different treatment is because the $\Td\geq 20\,$K 
subsample only represents 12\% of the high signal-to-noise points in the
observed data, but represents 28\% of those points in the simulated data. 
Therefore, modeling the entire sample of observed data points yields results
that are nearly identical to modeling only the $\Td < 20\,$K subsample, because
these points are the majority of data points.  For the simulated data, this is 
not entirely the case,  because the $\Td\geq 20\,$K subsample is not such a 
negligible fraction of the complete sample; therefore completely separating the 
two subsamples was more important for the simulated data than for the observed 
data. 

Now the model results are compared with the inputs. The model-derived $\DT$
values are all systematically lower than the input $\DT$ values.  For the $\NDv$ 
and n(H$_2$) parameters,
we must find the corresponding parameters in the two-component, two-subsample 
models (because these were the models used to generate the simulated maps).  
The continuum emission of the majority of points in the $\Td<20\,$K subsample are 
dominated by the emission of component~0 and the continuum emission of all of the 
points in the $\Td\geq 20\,$K subsample are dominated by the emission of component~1.
{\it However,\/} since the parameters we are discussing are largely physical 
parameters of the molecular gas, the $\cOone$ line emission is a better guide in 
determining which component is the more relevant.  The $\cOone$ line emission of 
component~1 dominates that of component~0 for {\it all\/} the points, except for the 
small minority of points where the component~1 temperature is less than about 4$\,$K. 
Therefore, the $\NDv$ and n(H$_2$) values of the one-component models are identified
with the $\nvco$ and $n_{c1}$ values of the two-component models for both subsamples.
The resultant $\NDv$ values compare very favorably with the known input values:
the range of $\NDv$ values includes the input values of $\nvco$ for the $\Td<20\,$K
and the $\Td\geq 20\,$K subsamples.  Also four of the five $\NDv$ values for the 
different SF values are within a factor of 2 of the input value for the 
$\Td\geq 20\,$K subsample.  The densities determined from the 
one-component models cover ranges that include the input values.  At SF=1.0, the 
model density is within a factor of 2 of the input $n_{c1}$ value of the 
$\Td\geq 20\,$K subsample.  Even though the one-component model curves do not 
characterize the data well, it is ironic that some parameter values, like the 
column density per velocity interval and the volume density, are obtained more 
accurately with the one-component, two-subsample models than with the simple 
two-component models.  It is clear then that, for some parameters, there is a 
greater advantage in having two subsamples than there is in having two components.
This may be an effect of using the continuum emission, because the two subsamples 
{\it almost\/} correspond to the two separate components when we consider just the
continuum emission.       

\begin{figure}
\epsscale{0.7}
\plotone{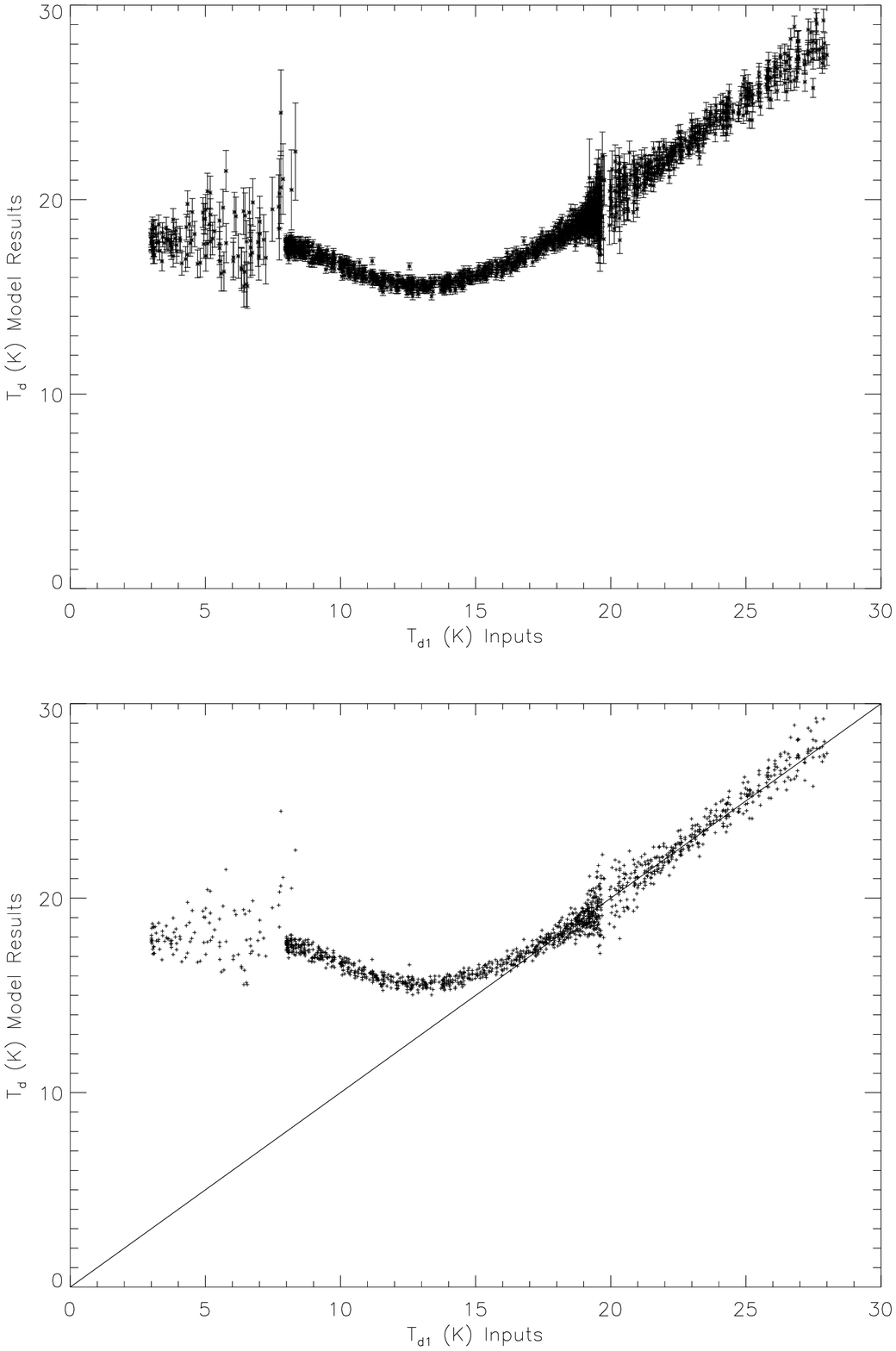}
\caption{The dust temperature as derived from fitting a one-component model is 
plotted against the component-1 dust temperature input values for the simulated 
data.  The upper panel includes the error bars in the model results, while the 
lower panel omits these error bars.  The lower panel also includes a solid 
straight line that represents $\Tdo(model)=\Tdo(input)$ for comparison with the 
plotted points.  The plots only include those pixels with the intensities above the 
5-$\sigma$ level in $\Ia$, $\Ib$, $\Ic$ {\it simultaneously\/}. 
\label{fig52}}
\end{figure}

    The best fitting model curves in Figure~\ref{fig50} were used to find 
the dust temperatures and the column densities as a function of position. 
Figure~\ref{fig52} shows the recovered $\Td$ values plotted against the input 
$\Tdo$ values.  As expected, the model $\Td$ values do not reproduce all the 
input $\Tdo$ values, except for high $\Tdo$.  Above $\Tdo\simeq 16\,$K, the 
model $\Td$ values are within about 1$\,$K of the input $\Tdo$ values.  Below
this temperature, the one-component $\Td$ values increase with {\it de}creasing
$\Tdo$.  This is because, as $\Tdo$ decreases, component~0 and its temperature
increasingly dominate the emission.  Also visible are two areas of larger error bars 
and, consequently, of heightened noise (i.e. more scatter), located at $\Tdo\lsim 
7\,$K and at $\Tdo\simeq 19$-21$\,$K.  This is due to the relatively larger noise 
at these temperatures in all three simulated maps (i.e., the maps of $\Ia$, $\Ib$, 
and $\Ic$). 

\begin{figure}
\epsscale{0.65}
\plotone{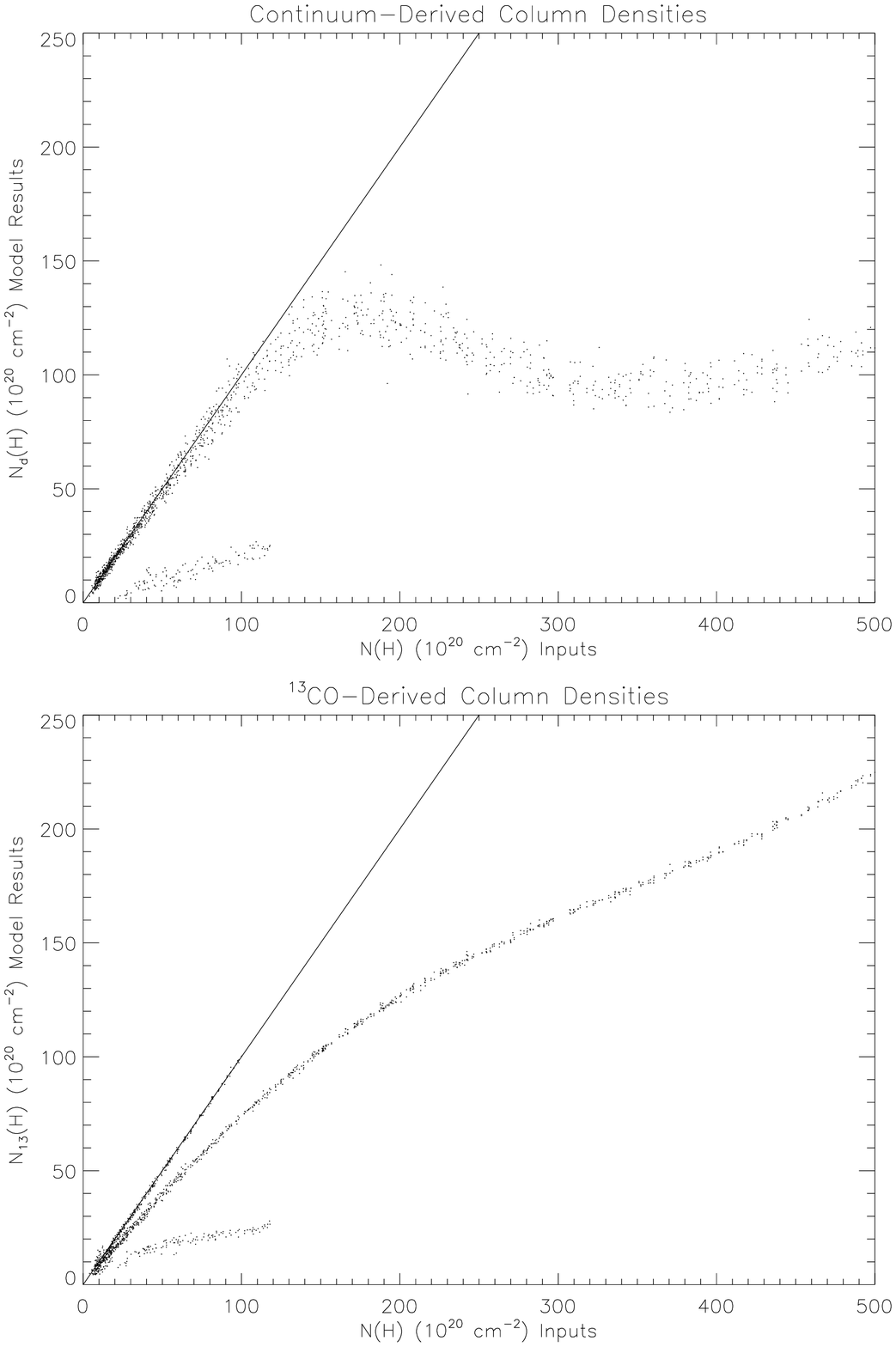}
\caption{Plots of the model gas column densities, $\NHd$ and $\Nnth$, for the 
one-component case versus the input column density values are shown for the 
simulations.  All column densities are in units of $10^{20}\ H\ nuclei\cdot\unit cm^{-2}$.  
The upper panel is the plot for the continuum-derived column densities, 
$\NHd$, and the lower panel is for the $\cO$-derived column densities, 
$\Nnth$. The panels omit the error bars for clarity.  A solid line is included
in each that represents the hypothetical case of agreement between the inputs and 
the model results (i.e. slope=1 and y-intercept=0).  The panels only include those 
pixels with intensities above the 5-$\sigma$ level in $\Ia$, $\Ib$, $\Ic$ 
{\it simultaneously\/}.   
\label{fig53}}
\end{figure}

    The incorrectly determined $\Tdo$ values adversely affect the determination
of the column densities.  This is obvious in the panels of Figure~\ref{fig53},
analogous to the plots in the previous subsections.  To explain these plots we 
consider three groups of points defined in terms of the plots 
that appear in Figure~\ref{fig40}.  The separate group of points that occur 
between $\Tdo = 3$ and 7$\,$K for all values of $\NHd$ and for $\Tdo$ between 
about 7 and 17$\,$K for $\NHd\lsim 50$ will be called ``Group~1".  The long 
descending (as one moves left to right) curve of points that starts at 
$\Tdo\simeq 7$ or 9$\,$K (for the simulations and observations, respectively) 
with $\NHd = 550$ and runs down to $\Tdo\simeq 20\,$K with $\NHd\simeq 10$ will 
be called ``Group~2".  The final ascending curve of points beyond $\Tdo\simeq 
20\,$K is ``Group~3".   In the panels of Figure~\ref{fig53}, Group~1 is the 
lower spur of points that runs from about (0,0) to about (100,20).  This spur 
corresponds to the lower spur in the column density plots of
Figures~\ref{fig43} and \ref{fig48}; as explained previously, the strong 
underestimates of the column densities represented by this spur is due to the 
strong overestimates of the dust temperatures: component-0 emission overwhelmingly
dominates over component-1 emission when this latter component is so cold.  
Assuming a single component in the modeling will then result in a dust 
temperature that is the component-0 dust temperature.  For the $\cO$-derived
column densities, Group~3 is the group of points that runs along the slope=1
line.  The nearly perfect agreement here is because the dust temperatures
for this group are correct to within a fraction of a Kelvin.  Group~2 is the
long curve that runs from the origin to the upper right of the plot in the
panels for the $\cO$-derived column densities.  As one ascends this curve
(moving from left to right), the column density estimates move increasingly
further from the correct (i.e., input) column densities.  In 
Figure~\ref{fig52}, Group~2 is the flattened V-shaped curve of points
that extends from $\Tdo\simeq 7\,$K to 20$\,$K.  As one moves to lower $\Tdo$,
the model $\Td$ moves further from the input $\Tdo$.  And, as one moves to
lower $\Tdo$ in Figure~\ref{fig52}, one is moving to higher $\Nnth$ in 
Figure~\ref{fig53}.  Consequently, the model $\Td$ moving further from the
input $\Tdo$ is the reason that the model $\Nnth$ moves further from the
input N(H).  Another noticeable characteristic of the curve (of the points
in Group~2) in the lower panels of Figure~\ref{fig53} is that its slope 
increasingly deviates from unity when moving left to right and then, for 
input $\rm N(H)\gsim 360$, the slope curves back in the direction of slope=1.
This is simply a reflection of the flattened V-shaped curve
in Figure~\ref{fig52} when moving right to left.  The upper panels of 
Figure~\ref{fig53}, which have the continuum-derived column densities, show
more extreme deviations of the one-component-model column densities from the
input column densities.  These panels also show qualitatively similar, but 
more extreme, slope variations in the Group~2 points than in the lower panels.
This is because these continuum observations, at wavelengths close to the
Wien limit, are much more sensitive to errors in the temperature estimates.
Note also that the Group~2 and Group~3 points are blended in the upper panels
for $\NHd\lsim 100$ because of the higher uncertainties of some of the 140$\um$ 
observations compared with some of the $\cO$ observations.  Because of the
greater sensitivity of the continuum observations to errors in temperature,
the error in the estimated total gas mass is further from the correct value
than that estimated from the $\cO$ observations: the simulated continuum 
observations underestimate the total mass by 48\% and the simulated $\cO$
observations underestimate this mass by 40\%. 

\begin{figure}
\epsscale{0.67}
\plotone{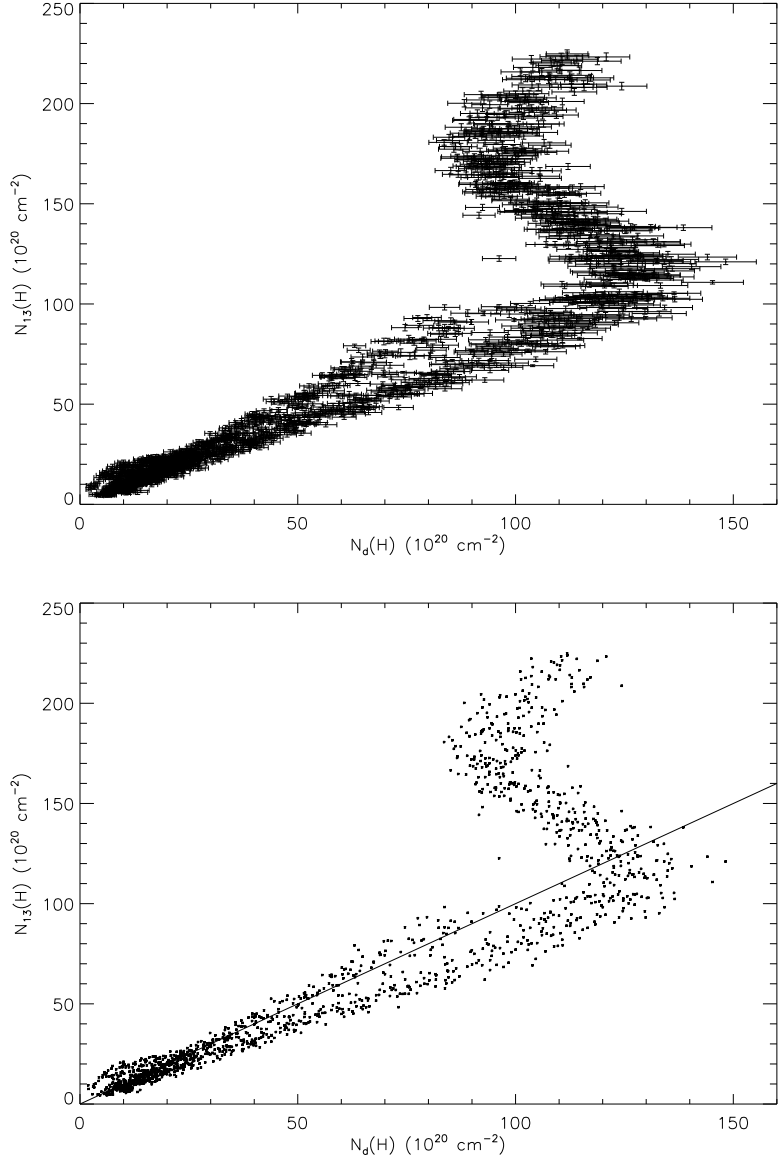}
\caption{Plots of the continuum-derived gas column densities, $\NHd$, versus
the $\cO$-derived gas column densities, $\Nnth$, are shown for the simulations,
where the column densities were derived using the parameters from the best-fit
one-component models.  All column densities are in units of $10^{20}\ H\ nuclei 
\cdot\unit cm^{-2}$.  The upper panel includes the error bars in the model results, while 
the lower panel omits these error bars.  The lower panel also includes a solid 
straight line that represents $\Nnth=\NHd$ for comparison with the plotted 
points.  The plots only include those pixels with the intensities above the 
5-$\sigma$ level in $\Ia$, $\Ib$, $\Ic$ {\it simultaneously\/}. 
\label{fig54}}
\end{figure}

Figure~\ref{fig54} has the plots of the $\cO$-derived column densities versus
the continuum-derived column densities.  In these plots, the disagreement is
no worse than a factor of 2 to within about 5\% for the majority of points
with $\NHd\gsim 10$.  The overall shape of the points, roughly reminiscent
of the Loch-Ness monster, roughly reflects the points in the upper
panels of Figure~\ref{fig53} about a solid line with slope=1 and intercept=0.
A better description is that the points in Figure~\ref{fig54} 
represent a reflection of the points in the upper panels of Figure~\ref{fig53} 
about the corresponding groups of points in the lower panels of that figure.
The slopes represented in Figure~\ref{fig54}, again for the points with 
$\NHd\gsim 10$, range between about 0.8 and about 2.  For the one-component
models applied to the real data, the slopes range from about 0.6 and 1.7
(see Figure~10 of Paper~I). 
 
    In summary, the one-component models can provide reasonable estimates
of the column density per velocity interval and volume density (i.e., within
factors of 2 or 3) provided that these models are applied to the two different 
subsamples (i.e. with $\Td$ below and above 20$\,$K); these reasonable numerical 
estimates are possible despite the poor characterization of the $\rd$ versus 
$\Td$ data points by the one-component models.  The estimates of $\DT$, however, 
can be wrong by about 20$\,$K.   The one-component models result in mass estimates 
that are too low by about 40-50\%; the continuum-derived mass estimates being worse 
on average than the $\cO$-derived mass estimates due to the higher temperature 
sensitivity of the continuum observations.

\section{Summary and Discussion\label{sec4}}

The reliability of recovering physical conditions in the dust and gas of molecular 
clouds using the far-IR continuum and the $\cOone$ line was tested by using simulated 
data.  These data were created using input beam-average column density and dust 
temperature maps that crudely represented the inferred physical conditions in the 
Orion~A and B giant molecular clouds \citep[see Paper~I][]{W05}.  Input physical 
parameters, with values similar to those recovered from modeling the actual observed 
data (see Paper~I), in combination with the column density and dust temperature maps 
gave us the simulated intensity maps in the 140$\um$ continuum, 240$\um$ continuum and 
$\cOone$ spectral line.  The simulated maps assumed two subsamples of positions within 
the clouds and two components.  The two components were component 0, with constant physical 
conditions within each subsample, and component 1, with constant physical conditions
within each subsample, except for spatially varying dust and gas temperatures.  The
two subsamples were defined by the component-1 dust temperature, $\Tdo$: those positions 
with $\Tdo<20\,$K represent one subsample and the positions with $\Tdo\ge 20\,$K 
represent the other subsample.  The point of the current paper was to apply the models
used in Paper~I to the simulated maps to see how well those models recover the input
values of the physical parameters.

Given that the simulated maps are based on the two-component, two-subsample models,
fitting such models to the simulated data in the noise-free case {\it might\/} be
expected to recover the inputs perfectly.  However, even in the noise-free case
some input parameters could {\it not\/} be recovered.  The component-0 and component-1
densities, for example, were an order of magnitude or more different from the inputs.
The simulated maps with noise show us that we can obtain the dust-gas temperature
difference, $\DT$, to within 1 or 2$\,$K {\it regardless of the specific value of 
$\DT$.\/}  The component zero dust temperature is apparently recovered to within 
a fraction of a Kelvin, but see Paper~III for further discussion of this.  Recovery of 
the component-1 dust temperatures and the gas column densities is accurate to within a 
few percent for 93\% of the points.  The other 7\% of the points have column densities 
too high or too low by a factor of about 5.  This results in overestimate of only 6-7\% 
in the total mass.

The simple two-component models applied to the simulations has shown what biases can
exist in the model results:
\begin{enumerate}
\item There are noticeable systematic offsets in the derived component-1 dust temperatures 
and in the derived column densities from their inputs.  These offsets come from forcing a 
single model curve to fit through the two different subsamples.
\item About 7\% of the column densities are wrong by factors of 5, as is the case for the 
two-subsample, two-component models.  Inspite of these systematic errors, the simple
two-component models overestimate the total mass of the clouds by only 3 to 6\%. 
\item Despite the varying the scale factors, the inferred component-1 densities are 
{\it all\/} systematically too low by an order of magnitude or more from the input density. 
\end{enumerate}
Keeping these shortcomings in mind gives us reasonable estimates of the parameter value 
ranges for all the two-component models as applied to the real observations.  These ranges 
are listed in Table~\ref{tbl-8}.  The range for the component-1 density is the kind of
range roughly expected for LTE emission of the $\cOone$ line.  The range for the component-1
column density per velocity interval is given, as stated earlier, by the large-scale cloud
properties at the low end and by the necessity of optically thin $\cOone$ emission at the
high end.  For component 0, the lower limit is nearly that of the master search grid. (Note 
that the $c_0\nvtcz$ product lower limit is indeed {\it equal\/} to that of the master search 
grid, but the $\nvtcz$ {\it itself\/} is still slightly larger than that.)
 
Fitting the one-component models to the simulated data shows that these models can provide 
reasonable estimates of the column density per velocity interval and volume density (i.e., 
within factors of 2 or 3) provided that these models are applied to the two different 
subsamples (i.e. with $\Td$ below and above 20$\,$K). The estimates of $\DT$, however, 
can be wrong by about 20$\,$K.   The one-component models result in mass estimates that 
are too low by about 40-50\%; the continuum-derived mass estimates being worse on average 
than the $\cO$-derived mass estimates due to the higher temperature sensitivity of the 
continuum observations.

These simulations have provided important insights into the reliability of the model results.
Yet other questions need to be addressed:
\begin{itemize}
\item What is the effect of the background subtractions used?  
\item How will dust associated with HI affect the results?  
\item Does changing the spectral emissivity index $\beta$ appreciably affect the results?  
\item Are there alternative kinds of models that would also explain the data?  
\item How representative are the results of the clouds as a whole, given that the modeled cloud 
positions only represent 26\% of the area of the Orion clouds ? 
\end{itemize}
Paper~III examines these questions and discusses the scientific implications of the results.

%% The \notetoeditor{TEXT} command allows the author to communicate
%% information to the copy editor.  This information will appear as a
%% footnote on the printed copy for the manuscript style file.  Nothing will
%% appear on the printed copy if the preprint or
%% preprint2 style files are used.

%% The eqnarray environment produces multi-line display math. The end of
%% each line is marked with a \\. Lines will be numbered unless the \\
%% is preceded by a \nonumber command.
%% Alignment points are marked by ampersands (&). There should be two
%% ampersands (&) per line.

%% Putting eqnarrays or equations inside the mathletters environment groups
%% the enclosed equations by letter. For instance, the eqnarray below, instead
%% of being numbered, say, (4) and (5), would be numbered (4a) and (4b).
%% LaTeX the paper and look at the output to see the results.


%% Included in this acknowledgments section are examples of the
%% AASTeX hypertext markup commands. Use \url without the optional [HREF]
%% argument when you want to print the url directly in the text. Otherwise,
%% use either \url or \anchor, with the HREF as the first argument and the
%% text to be printed in the second.

\acknowledgments This work was supported by CONACyT grants \#211290-5-0008PE and 
\#202-PY.44676 to W.~F.~W. at {\it INAOE.\/} I am very grateful to W.~T.~Reach
for his comments and support.  I owe a great debt of thanks to Y.~Fukui and 
T.~Nagahama of Nagoya University for supplying us with the $\cO$ data that made 
this work possible. The author is grateful to R.~Maddalena and T.~Dame, who 
supplied the map of the peak $\COone$ line strengths and provided important 
calibration information.  I thank D.~H.~Hughes, W.~T.~Reach, Y. Fukui, M.~Greenberg, 
T.~A.~.D.~Paglione, G.~MacLeod, E.~Vazquez Semadeni, and others for stimulating 
and useful discussions.

\clearpage


\clearpage 

\clearpage 

\clearpage 

\clearpage 

\clearpage 

\clearpage 

\clearpage 

\clearpage 

\clearpage 

\clearpage 

\clearpage 

\clearpage 

\clearpage 

\clearpage 

\clearpage 

\clearpage 

\clearpage 

\clearpage 

\clearpage

%% If you are not including electronic art with your submission, you may
%% mark up your captions using the \figcaption command. See the 
%% No more than seven \figcaption commands are allowed per page, 
%% so if you have more than seven captions, insert a \clearpage 

%% Tables should be submitted one per page, so put a \clearpage before

%% Two options are available to the author for producing tables:  the
%% deluxetable environment provided by the AASTeX package or the LaTeX
%% table environment.  Use of deluxetable is preferred.
%%

%% Three table samples follow, two marked up in the deluxetable environment,
%% one marked up as a LaTeX table.

%% In this first example, note that the \tabletypesize{}
%% command has been used to reduce the font size of the table.
%% Note also that the \label command needs to be placed 

\begin{deluxetable}{ccccccc}
%\tabletypesize{\scriptsize}
\tablecaption{Parameter Values for the Simulations\label{tbl-7}}
\tablewidth{0pt}
\tablehead{
\colhead{Parameter} & \colhead{Input Values}\span\omit
& \colhead{Model Results (Noise Free)}\span\omit 
& \colhead{Model Results (with Noise)}\span\omit\\
\colhead{} & \colhead{$\Tdc<20\,K$} & \colhead{$\Tdc\geq 20\,K$}
& \colhead{$\Tdc<20\,K$\tablenotemark{a}} 
& \colhead{$\Tdc\geq 20\,K$\tablenotemark{b}}
& \colhead{$\Tdc<20\,K$\tablenotemark{c}} 
& \colhead{$\Tdc\geq 20\,K$\tablenotemark{d}}
}
\startdata
$\DT$\tablenotemark{e} & 0 & 0 & $-1$ & 0 & $-1$ & 2 \\
\noalign{\bigskip}
$c_0$\tablenotemark{f} & 1.0 & 0.4 & 0.063 & 0.63 & 1.6 & 0.13 \\
\noalign{\medskip}
$\Tdz$\tablenotemark{e} & 18 & 18 & 18 & 18 & 18 & 18 \\
\noalign{\medskip}
$\nvtcz$\tablenotemark{g} & $5.0\times 10^{15}$ & $5.0\times 10^{14}$ 
& $5.0\times 10^{16}$ & $3.2\times 10^{14}$ 
& $2.0\times 10^{15}$ & $5.0\times 10^{15}$ \\
\noalign{\medskip}
$n_{c0}$\tablenotemark{h} & $3.2\times 10^{4\phantom{5}}$ & $1.0\times 10^{4\phantom{5}}$ 
& $5.6\times 10^{1\phantom{5}}$ & $1.0\times 10^{4\phantom{5}}$ 
& $3.2\times 10^{1\phantom{5}}$ & $1.0\times 10^{4\phantom{5}}$ \\
\noalign{\medskip}
$c_0\nvtcz$\tablenotemark{g} & $5.0\times 10^{15}$ & $2.0\times 10^{14}$ 
& $3.2\times 10^{15}$ & $2.0\times 10^{14}$ 
& $3.2\times 10^{15}$ & $6.5\times 10^{14}$ \\
\noalign{\bigskip}
$\nvtco$\tablenotemark{g} & $2.0\times 10^{16}$ & $5.0\times 10^{15}$  
& $1.3\times 10^{16}$ & $5.0\times 10^{15}$ 
& $1.3\times 10^{16}$ & $3.2\times 10^{15}$ \\
\noalign{\medskip}
$n_{c1}\tablenotemark{h}$ & $3.2\times 10^{4\phantom{5}}$ & $5.6\times 10^{3\phantom{5}}$ 
& $5.6\times 10^{3\phantom{5}}$ & $5.6\times 10^{3\phantom{5}}$ 
& $5.6\times 10^{3\phantom{5}}$ & $1.0\times 10^{4\phantom{5}}$ \\
\noalign{\bigskip}
$\chi_\nu^2$ & --- & --- & $1.59\times 10^{-2}$ 
& $6.23\times 10^{-4}$ & 1.15 & 1.90 \\
\noalign{\medskip}
$\nu$ & --- & --- & 1129 & 366 & 1066 & 389 \\
\enddata

\tablenotetext{a}{Formal relative errors are $\leq 1\times 10^{-5}$ for all
parameters, except $\DT$ and $\Tdz$, which have formal absolute errors of 
$\leq 1\times 10^{-5}\unit K$.}
\tablenotetext{b}{Formal relative errors are $\leq 1\times 10^{-1}$ for all
parameters, except $\DT$ and $\Tdz$. $\DT$ has a formal absolute error of 
$\leq 1\times 10^{-1}\unit K$.  $\Tdz$ was simply adopted to be 18$\,$K.}
\tablenotetext{c}{Formal relative errors are $\leq 3\times 10^{-5}$ for all
parameters, except $\DT$ and $\Tdz$, which have formal absolute errors of 
$\leq 3\times 10^{-5}\unit K$.}
\tablenotetext{d}{Formal relative errors are $\leq 2\times 10^{-2}$ for all
parameters, except $\DT$ and $\Tdz$.  $\DT$ has a formal absolute error of 
$\leq 2\times 10^{-2}\unit K$.  $\Tdz$ was simply adopted to be 18$\,$K.}
\tablenotetext{e}{In units of Kelvins.}
\tablenotetext{f}{Dimensionless.}
\tablenotetext{g}{In units of $\cO$ molecules$\,\cdot\ckms$.}
\tablenotetext{h}{In units of H$_2$ molecules$\,\cdot\unit cm^{-3}$.}

%\tablecomments{}

\end{deluxetable}

\clearpage

\begin{deluxetable}{cc}
%\tabletypesize{\scriptsize}
\tablecaption{Best Estimates of Parameter Value Ranges\tablenotemark{a}
\label{tbl-8}}
\tablewidth{0pt}
\tablehead{
\colhead{Parameter} & \colhead{Range of Values}
}
\startdata
$\DT$\tablenotemark{b} & $-$1 to +2$\,$K \\
\noalign{\medskip}
%%$\Tdz$ & 18$\,$K\tablenotemark{c}\\
$\Tdz$ & 18$\,$K\tablenotemark{c}\\
\noalign{\bigskip}
$c_0\nvtcz$ & $2.0\times 10^{14}$ to $5.0\times 10^{15}\ \cOit\ \ckms$\\
\noalign{\medskip}
$n_{c0}$ & $\gsim 20\ H_2\rm\ cm^{-3}$\\
\noalign{\bigskip}
$\nvtco$\tablenotemark{d} & $3\times 10^{15}$ to $2\times 10^{16}\ \cOit\ \ckms$ \\
\noalign{\medskip}
$n_{c1}$ & $\gsim few\times 10^3\ H_2\rm\ cm^{-3}$ \\
\enddata

\tablenotetext{a}{See Subsection~\ref{sssec372} for details.}
\tablenotetext{b}{Assuming two-component models applied to {\it both\/} subsamples.}
%%\tablenotetext{c}{The range here is difficult to determine. See the assumptions 
%%mentioned in Section~\ref{sssec372}.}
\tablenotetext{c}{The uncertainty of this will be dealt with in Paper~III.}
\tablenotetext{d}{For the two-component models applied to the two subsamples,
the $\nvtco$ value would be at the higher end of this range for the $\Tdc < 20\,$K
subsample and at the lower end for the $\Tdc\geq 20\,$K subsample.}

%\tablecomments{}

\end{deluxetable}

\clearpage

\end{document}